\documentstyle[12pt,psfig,graphicx,epsfig,preprint,aps,tighten]{revtex}
\def\baselinestretch{1.24}
\begin{document}
\draft

\newcommand{\be}{\begin{equation}}
\newcommand{\ee}{\end{equation}}
\newcommand{\lsim}   {\mathrel{\mathop{\kern 0pt \rlap
  {\raise.2ex\hbox{$<$}}}
  \lower.9ex\hbox{\kern-.190em $\sim$}}}
\newcommand{\gsim}   {\mathrel{\mathop{\kern 0pt \rlap
  {\raise.2ex\hbox{$>$}}}
  \lower.9ex\hbox{\kern-.190em $\sim$}}}
\def\be{\begin{equation}}
\def\ee{\end{equation}}
\def\ba{\begin{eqnarray}}
\def\ea{\end{eqnarray}}

\def\i{{\rm i}}
\def\d{{\rm d}}
\def\e{{\rm e}}

\def\ap{\approx}
\def\F{{\cal F}}
\def\G{{\cal G}}

\def\PLZ{P_{\rm LSZ}}
\def\Ee{\langle E_{\bar\nu_{\rm e}}\rangle}
\def\Eh{\langle E_{\bar\nu_{\rm h}}\rangle}
\def\Eb{E_{\rm b}}

\def\t{\vartheta}
\def\tm{\vartheta_{\rm m}}
\def\Dm{\Delta_{\rm m}}

\font\menge=bbold9 scaled \magstep1
\def\nota#1{\hbox{$#1\textfont1=\menge $}}
\def\R{\nota R}

\title{
              {\small CERN-TH/2001-219}\hfill
              {\small IFIC/01-42}\hfill
              {\small IFUP--TH/2001-25}\hfill
              {\small MPI-PhT/2002-23}\\          
 \vskip1.0cm  SN~1987A and the Status of Oscillation Solutions\\
 to  the Solar Neutrino Problem\\
 {\small (including an appendix discussing the NC and day/night data
 from SNO)}}

\author{M. Kachelrie{\ss}$^{1,2}$, A. Strumia$^1$\footnote{On leave from
dipartimento di Fisica dell'Universit\`a di Pisa
and INFN.}, R. Tom{\`a}s$^{2,3}$ and J.W.F. Valle$^3$}

\address{$^1$TH Division, CERN, CH-1211 Geneva 23, Switzerland}
\address{$^2$Max-Planck-Institut f\"ur Physik, D-80805 M\"unchen, Germany}
\address{$^3$Instituto de Fisica Corpuscular -- C.S.I.C.
          -- Universitat de Val{\`e}ncia  \\
          Edificio de Institutos   -- Apartado de Correos 22085 -
          46071 Val{\`e}ncia, Spain}

\date{\today}

\maketitle

\begin{abstract}
  We study neutrino oscillations and the level-crossing probability
  $\PLZ$ in power-law potential profiles $A(r)\propto r^n$. We give
local
  and global adiabaticity conditions valid for all mixing angles $\t$
  and discuss different representations for $\PLZ$.  For the
  $1/r^3$ profile typical of supernova envelopes we compare our
  analytical to numerical results and to earlier approximations used
  in the literature. We then perform a combined likelihood analysis of
  the observed SN~1987A neutrino signal and of the latest solar
  neutrino data, including the recent SNO CC measurement.  We find
  that, unless all relevant supernova parameters (released binding
  energy, $\bar\nu_e$ and $\bar\nu_{\mu,\tau}$ temperatures) are near
their
  lowest values found in simulations, the status of large mixing type
  solutions deteriorates considerably compared to fits using only
  solar data.  This is sufficient to rule out the vacuum-type
  solutions for most reasonable choices of astrophysics parameters.
  The LOW solution may still be acceptable, but becomes worse than the
  SMA--MSW solution which may, in some cases, be the best combined
  solution.  On the other hand the LMA--MSW solution can easily
  survive as the best overall solution, although its size is generally
  reduced when compared to fits to the solar data only.
\end{abstract}


\def\baselinestretch{1.6}

\section{Introduction}
\label{sec:introduction}

The detection of astrophysical neutrinos has played a major r{\^o}le in
the still on-going process of establishing neutrino masses and
mixing~\cite{review}.  The propagation of these neutrinos from their
source to the detector can be influenced by matter effects as it was
pointed out by Wolfenstein~\cite{Wo78}. In particular, Mikheev and
Smirnov~\cite{Mi85} showed that the flavour of neutrinos produced in
the core of the sun or a supernova can be efficiently converted into a
different one even for small mixing angles $\t$ via ``resonant''
neutrino oscillations (MSW effect). In the limit that the density
inside a star varies much slower than the typical distance
characterizing flavour oscillations, the instantaneous matter states
of the neutrino propagating towards the vacuum change adiabatically
their flavour composition which can be reversed completely even for
small $\t$.

The analytical study of non-adiabatic neutrino oscillations started
soon after the discovery of the MSW effect. The leading non-adiabatic
effects were calculated for a linear potential profile in
Ref.~\cite{cross} as a Landau-St{\"u}ckelberg-Zener (LSZ) crossing
probability~\cite{LSZ},
\be
 P_{\rm LSZ} =  \exp \left( -\frac{\gamma\pi}{2} \right) \,.
\ee
The adiabaticity parameter for a linear profile is
\be  \label{gamma0}
 \gamma =   \frac{|m_2^2-m_1^2|\, \sin^22\t}
                 {2E\cos 2\t \:|\d\ln A/\d x|_0} \,,
\ee
where $E$ is the energy, $m_i$ denotes the masses, and $\t$ the vacuum
mixing angle of the two (active) neutrinos. Furthermore,
$A=2EV=2\sqrt{2}G_F N_e E$ is the induced mass squared for the
electron neutrino. The parameter $\gamma$ has to be evaluated at the
so-called resonance point, i.e. the point where the mixing angle in
matter is $\tm=\pi/4$. For a linear profile, adiabaticity is maximally
violated at this point.  Therefore, the probability that a neutrino
jumps from one branch of the dispersion relation to the other one is
indeed maximal at the resonance. Since the condition $\tm=\pi/4$ can be
fulfilled only for a normal mass hierarchy in the case of neutrinos
and for an inverted hierarchy in the case of anti-neutrinos,
Eq.~(\ref{gamma0}) allows the calculation of $\PLZ$ only in half of
the parameter space of neutrino mixing.

Later, Kuo and Pantaleone \cite{Ku89} derived the LSZ crossing
probability for an arbitrary power-law like $r^n$ potential profile.
This type of profile does not only contain the case $n \ap -3$ typical
for supernova envelopes, but also the exponential profile of the sun
in the limit $n\to\pm\infty$. Moreover, it allows discussing which
features of
neutrino oscillations are generic and which ones are specific for,
e.g., the linear profile $n=1$ usually discussed. Kuo and Pantaleone
found that also for arbitrary power-law like potential profiles the
dependence on the neutrino masses and energies can be factored out,
while the effect of a non-linear profile can be encoded into a
correction function $\F_n$,
\be \label{WKB0}
 \PLZ =  \exp \left( -\frac{\pi\gamma_n}{2} \:  \F_n(\t) \right) \,,
\ee
which only depends on $\t$ and $n$. The ``adiabaticity'' parameter
$\gamma_n$ has to be evaluated still at the resonance although, as we
will
show, it does not coincide with the point of maximal violation of
adiabaticity (PMVA) for $n\neq 1$.  An unsatisfactory feature of
Eq.~(\ref{WKB0}) is its restricted range of applicability: As in the
case of a linear profile, the calculation of $\PLZ$ with
Eq.~(\ref{WKB0}) requires the validity of the resonance condition
$\tm=\pi/4$. Several other authors have also considered non-adiabatic
effects in neutrino oscillations, partially for arbitrary potential
profiles, but their results either do not allow simple numerical
evaluation or have also a restricted range of validity~\cite{others}.
Therefore, it has not been possible to calculate, e.g., the survival
probability of supernova anti-neutrinos in the quasi-adiabatic regime
without solving numerically their Schr{\"o}dinger equation.

The purpose of this work is twofold\footnote{Some of the results of
  this work have been already briefly presented in Ref.~\cite{l}.}.
First, we clarify the physical significance of the resonance point
compared to the point where adiabaticity is maximally violated {\em
  locally\/}: we find that the crossing probability has its maximum at
the PMVA and not at $\tm=\pi/4$.  Then we show explicitly that the
product $\gamma_n\F_n$ can be evaluated at an arbitrary point.  We
conclude
that the ``resonance'' point has in general no particular physical
meaning: it does not necessarily describe the point of maximal
violation of adiabaticity nor is it necessary to calculate the
adiabaticity parameter at the resonance.  We provide a criterion that
measures the {\em global\/} cumulative non-adiabatic effects along the
neutrino trajectory and gives an estimate of the border between the
adiabatic and non-adiabatic regions for a power-law profile.
Moreover, we obtain an accurate formula for $\PLZ$ which is valid for
all $\t$ and convenient for numerical evaluation.

Second, we apply this formula to neutrino oscillations in supernova
(SN) envelopes and in particular to the neutrino signal of SN~1987A.
Performing a combined likelihood analysis of the observed neutrino
signal of SN~1987A and the updated global set of solar neutrino data
including the recent SNO CC measurement~\cite{solar,solar2}, we find that
the supernova data offer additional discriminating power between the
different solutions.
We find that, unless all relevant supernova parameters (released
binding energy, $\bar\nu_e$ and $\bar\nu_{\mu,\tau}$ temperatures) are
near
their lowest values found in simulations, the status of the large
mixing solutions to the solar neutrino problem deteriorates
considerably compared to fits using only solar data.  This is
sufficient to rule out the vacuum-type solutions for most reasonable
choices of astrophysics parameters. The LOW solution may still be
acceptable, but becomes worse than the SMA--MSW solution.  In contrast
the LMA--MSW solution can easily survive as the best overall solution,
although its size is generally reduced when compared to fits to the
solar data only. In the analysis of the solar neutrino
data which we adopt in the present paper, the SMA--MSW solution is
absent at $3$ standard deviations if solar data only are included but
may reappear once
SN~1987A data are included and, may, in some cases, be the
best combined solution.


\section{Neutrino evolution: resonance  and adiabaticity conditions,
  maximal violation of adiabaticity}
\label{sec:neutr-evol-reson}

We consider neutrino oscillations in a two flavour scenario and
label always the heavier neutrino mass eigenstate with ``2''. Then
$\Delta$ is positive and the vacuum mixing angle $\t$
is in the range $[0\!:\!\pi/2]$. As starting point for our discussion,
we use the evolution equation for the medium states $\tilde\psi$ first
given in Ref.~\cite{Mi87},
\be
 \frac{\d}{\d r}
 \left( \begin{array}{c} \tilde\psi_1 \\ \tilde\psi_2 \end{array}
 \right) =
 \left( \begin{array}{cc} {\i}\Dm/(4E) & -\tm^\prime \\
                          \tm^\prime & -{\i}\Dm/(4E)  \end{array}
 \right)
 \left( \begin{array}{c} \tilde\psi_1 \\ \tilde\psi_2 \end{array}
 \right) \,.
\ee
Here,
\be
 \Dm = \sqrt{\left(A-\Delta\cos 2\t\right)^2 + (\Delta\sin 2\t)^2}
\ee
denotes the difference between the effective mass of the two neutrino
states in matter, $\tm$ is the mixing angle in matter with
\be
 \tan 2\tm = \frac{\Delta\sin 2\t}{\Delta\cos 2\t -A}
\ee
and $\tm^\prime=\d\tm/\d r$. 
Following Ref.~\cite{Fr00}, we rewrite the evolution
equation as
\be  \label{S}
 \frac{\d}{\d\tm}
 \left( \begin{array}{c} \tilde\psi_1 \\ \tilde\psi_2 \end{array}
 \right) =
 \left( \begin{array}{cc} {\i}\Dm/(4E\tm^\prime) & -1 \\
                          1 & -{\i}\Dm/(4E\tm^\prime)  \end{array}
 \right)
 \left( \begin{array}{c} \tilde\psi_1 \\ \tilde\psi_2 \end{array}
 \right)
\ee
with
\be \label{tm} 
 \tm^\prime = \frac{\sin^22\tm}{2\Delta\sin2\t} \:\frac{\d A}{\d r} \,,
\ee
\be \label{dt}
 \frac{\Dm}{\tm^\prime} = \frac{2\Delta^2\sin^22\t}{\sin^32\tm}\:
                          \frac{1}{\d A/ \d r} \,,
\ee
and
\be \label{A}
 A = \frac{\Delta\sin(2\tm-2\t)}{\sin 2\tm} \,.
\ee

The traditional condition for an adiabatic evolution of a neutrino
state along a certain trajectory is that the diagonal entries of the
Hamiltonian in Eq.~(\ref{S}) are large with respect to the
non-diagonal ones, $|\Dm|\gg |4E\tm^\prime|$. This condition measures
indeed
how strong adiabaticity is {\em locally\/} violated. Therefore, the
PMVA is given by the minimum of $\Dm/\tm^\prime$. Differentiating
Eq.~(\ref{dt}) for a power law profile, $A(r)\propto r^n$, we find the
minimum at
\ba  \label{eq_pmva}
 && \cot(2\tm-2\t) + 2\cot(2\tm)
- \frac{1}{n} \, \left[ \cot(2\tm-2\t) - \cot(2\tm) \right] = 0 \,.
\ea
For $n=1$, the PMVA is at $\tm=\pi/4$ for all
$\t$. Thus, in the region where the resonance point is
well-defined, it coincides with it.
In the general case, $n\neq 1$, the PMVA agrees however only for
$\t=0$ with the resonance point.
Finally, we recover the result of Ref.~\cite{Fr00} for an exponential
profile in the limit $n\to\pm\infty$.

\begin{center}
\begin{figure}
\epsfig{file=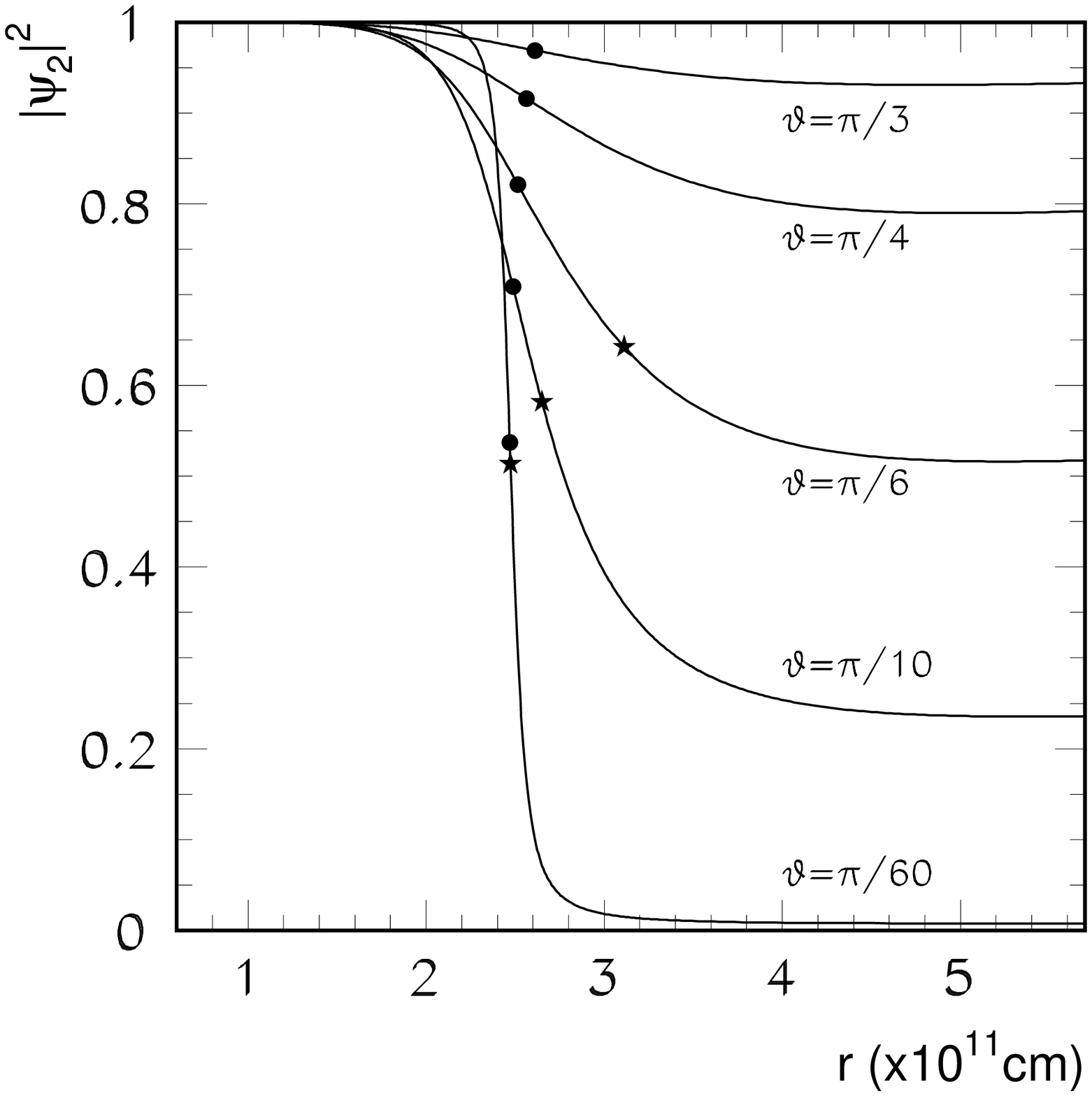,height=7.cm,width=9.cm,angle=0}
\caption{\label{pmva}
  Survival probability $p(r)=|\tilde\psi_2(r)|^2$ as function of $r$ for
  a neutrino produced at $r=0$ as $\tilde\nu_2$. The point of maximal
  violation of adiabaticity predicted by Eq.~(\ref{eq_pmva}) is
  indicated by a dot, while the resonance point for a power law
  profile $A=2EV_0 (r/R_0)^{-3}$ with $\Delta/E = 2\times
  10^{-10}~\mathrm{eV}^2/\mathrm{MeV}$, $V_0 = 1.5\times 10^{-9}$~eV and
  $R_0=10^9$~cm is denoted by a star.}
\end{figure}
\end{center}

\begin{center}
\begin{figure}
\epsfig{file=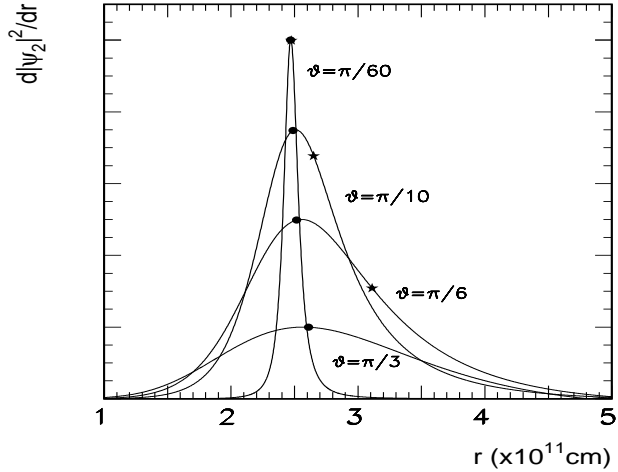,height=7.cm,width=9.cm,angle=0}
\caption{\label{pmva2}
  Change of the survival probability $\d p(r)/\d r$ of a neutrino
  produced at $r=0$ as $\tilde\nu_2$ together with the point of maximal
  violation of adiabaticity (dot) and the resonance point (star) for a
  power law profile $A\propto r^{-3}$. The height of the different
curves is
  rescaled.}
\end{figure}
\end{center}

In Fig.~\ref{pmva}, we show the survival probability
$p(r)=|\tilde\psi_2(r)|^2$ for a neutrino produced at $r=0$ as
$\tilde\nu_2$, together with the PMVA predicted by Eq.~(\ref{eq_pmva})
and the resonance point for a power law profile $A\propto r^{-3}$.  The
resonance condition predicts a transition in regions of density
lower than that which characterizes the PMVA, until for
$\t=\pi/4$ the resonance point reaches $r=\infty$ and the concept of a
resonant transition breaks down completely.  If one plots the change
of the survival probability, $\d p(r)/\d r=\d|\tilde\psi_2(r)|^2/\d r$,
as function of $r$, cf.~Fig.~\ref{pmva2}, it can be clearly seen that
Eq.~(\ref{eq_pmva}) describes quite accurately the most probable
position of the level crossing, while the resonance condition fails.
As an immediate consequence, we note that in the case that the true
potential profile $A(r)$ is only approximately given by a power-law,
its exponent should be determined by the region around the PMVA, not
by the region around the resonance point. Finally, Fig.~\ref{pmva2}
shows that the crossing probability becomes less and less localized
near the PMVA for larger mixing angles $\t$.

Let us now discuss the condition for the adiabatic evolution of a
neutrino state along a trajectory from the core of a star to the
vacuum. While the condition $|\Dm|\gg |4E\tm^\prime|$ indicates whether
adiabaticity is locally violated, we need now a global criterion that
measures the cumulative non-adiabatic effects along the trajectory
from $\tm\ap\pi/2$ to $\t$. For a non-adiabatic evolution of the
neutrino state we require that
\be
 \bigg| \!\int_{\pi/2}^\t\d\tm \: \tilde\psi_1  \bigg| = \varepsilon\:
  \bigg| \!\int_{\pi/2}^\t\d\tm \: \frac{4E\tm'}{\Dm} \: \tilde\psi_2
\bigg|
\ee
with $\varepsilon\ll 1$. Then we can use
$\tilde\psi_1\ap\cos\tm$, $\tilde\psi_2\ap\sin\tm$ and Eq.~(\ref{dt}),
\be
 \bigg| \!\int_{\pi/2}^\t\d\tm \:\cos\tm \bigg| =
 \frac{2E\varepsilon}{\Delta^2\sin^2(2\t)} \:
 \bigg|\!\int_{\pi/2}^\t\d\tm \sin\tm \sin^3 2\tm\frac{\d A}{\d r}\:
\bigg|  \,.
\label{RHS}
\ee

We consider first the simple case of an {\em exponential\/} profile,
$|\d A/\d r|= A/R_0$, where we can solve the $\tm$ integral of the RHS
of Eq.~(\ref{RHS}) analytically. Then the non-adiabatic region obeys
the following condition
\be  \label{crit}
 \frac{\Delta}{E} =
 \varepsilon\: \frac{f(\t)}{(1-\sin\t)\sin^2(2\t)} \: \frac{2}{R_0}
\ee
with
\be   \label{ft}
 f(\t)= \frac{1}{3}\sin\t    -\frac{1}{10} \sin(3\t)  +
        \frac{1}{42}\sin(5\t) +\frac{16}{35}\cos(2\t)  \,.
\ee
This criterion agrees well with the one derived in a more intuitive way
in
Ref.~\cite{Fr00},
\be  \label{fried}
 \frac{\Delta}{E} = \varepsilon'\:
\frac{\tan(\pi/4+\t/2)}{2\sin\t\tan\t} \:
                    \frac{1}{R_0} \,,
\ee
for $\varepsilon\ap 2\varepsilon'$.

In the case of a {\em power-law\/} like profile, $A(r)=2EV_0
(r/R_0)^n$, the border between the adiabatic and non-adiabatic regions
is given by
\be
\label{b}
\frac{\Delta}{E} = \left\{ \varepsilon\:
\frac{f(\t)}{\sin^2(2\t)(1-\sin\t)} \:
                  \frac{2n (2V_0)^{1/n}}{R_0}\right\}^{\frac{n}{n+1}}
\,,
\ee
with
\be
\label{i}
 f(\t) =  \bigg|\int_{\pi/2}^\t\d\tm \sin\tm \left[ \sin(2\tm)
\right]^{2+1/n}
\left[ \sin(2\tm-2\t) \right]^{1-1/n} \bigg|  \,.
\ee
In Fig.~\ref{border1} we compare the different predictions for the
borders between the non-adiabatic and adiabatic regions for solar
neutrinos with energy $E=1$~MeV with the contours of constant survival
probability (dashed lines) of the neutrino eigenstate $\tilde\nu_2$
obtained by solving the Schr{\"o}dinger equation~(\ref{S}). In the solar
case, we can approximate the potential by an exponential profile with
the typical solar height scale $R_0=R_\odot/10.54$.  The difference
between our general criterion~(\ref{crit}) together with (\ref{ft})
compared to (\ref{fried}) is negligible for $\varepsilon' =
\varepsilon/2=1$: both
criteria predict very well the borderline of the adiabatic region.
For comparison, the 90, 95, 99 and 99.73\% confidence level (C.L.)
contours for the different solutions to the solar neutrino problem
from Ref.~\cite{solar} are also shown.

In Fig.~\ref{border2} we show a similar comparison for anti-neutrinos
with energy $E=20$~MeV and a profile typical for supernova envelopes,
$V(r)= 1.5\times 10^{-9}$~eV $(10^9~{\rm cm}/r)^{3}$.  We now compare
the
borderlines obtained using our general criterion~(\ref{b}) together
with two different choices for $f(\t)$: the solid line is obtained
solving numerically (\ref{i}) for $n=-3$, while we used (\ref{ft}) for
the dash-dotted line. We find that already for moderate $|n|$ values
the function $f(\t)$ depends rather weakly on $n$ so that the
expression (\ref{ft}) obtained for $n\to\pm\infty$ can be used as a
reasonable
approximation already for $n=-3$.

\section{The crossing probability in the WKB formalism}
\label{sec:cross-prob-wkb}

\subsection{The correction functions $\F_n$}
\label{sec:corr-funct-f_n}

We discuss now in detail the calculation of the leading term to the
crossing probability within the WKB formalism. The semi-classical
transition probability $\PLZ$ from the neutrino states $1\to 2$ under an
adiabatic change of the action is mainly determined by those $x$
values for which $E_1(x)=E_2(x)$. Using the ultra-relativistic limit
and omitting an overall phase, the WKB formula gives
\be  \label{start}
 \ln\PLZ = -\frac{1}{E}\: \Im\int_{x_1(A_1)}^{x_2(A_2)} \d x \:
           \left[(A-\Delta C)^2 + (\Delta S)^2 \right]^{1/2} \,,
\ee
where $A_2=\Delta(C+{\i} S)=\Delta e^{2{\i}\t}$ is the branch point of
$\Dm$ in the
upper complex $x$ plane and  we have also introduced the
abbreviations $C=\cos 2\t$ and $S=\sin 2\t$.

We identify the physical coordinate $r\in [0\!:\!\infty]$ with the
positive part of the real $x$ axis, i.e. we consider a neutrino state
produced at small but positive $x$ propagating to $x=\infty$. Then, a
convenient choice for $A_1$ is to use the real part of $A_2$ for
$C>0$, i.e. the ``resonance'' point $A_1=\Delta C$. However, we stress
that this choice has technical reasons and makes sense only for $C>0$:
consider for instance the simplest case $n=1$. Then both the
integration path chosen and the branch cut are for $C<0$ in the
half-plane $\Re(x)<0$. The physical interpretation is therefore that
an anti-neutrino state created at small but negative $x$ propagates to
$-\infty$. This case is however equivalent to a neutrino state
propagating with $C>0$ in the right half-plane and therefore contains
no new information. Thus, we expect the correction functions $\F_n$
obtained with the integration path from $\Delta C$ to
$\Delta\e^{2{\i}\t}$ to be functions with period $\pi/4$ and to be valid
only in the resonant region.

We substitute first $A=\Delta(B+C)$,
\be   \label{app}
 \ln\PLZ = -\frac{\Delta}{E}\: \Im\int_{0}^{{\i} S}
           \d B \: \frac{\d x}{\d B} \: (B^2 + S^2 )^{1/2} \,.
\ee
Then we expand the Jacobian for a potential $A(x)=A_0(x/R_0)^n$,
\be
 f = \frac{\d x}{\d B} = \frac{R_0}{n}
 \left(\frac{\Delta}{A_0}\right)^{1/n} (B+C)^{1/n-1} \,,
\ee
into a power series around the arbitrary point $B_0$.  Using
\be
 \left. \frac{1}{A} \, \frac{\d A}{\d x} \, \right|_{B_0} =
 \left. \frac{n}{x} \, \right|_{B_0} =
 \frac{n}{R_0}  \left(\frac{A_0}{\Delta}\right)^{1/n}
 \frac{1}{(B_0+C)^{1/n}} \,,
\ee
we obtain
\be  \label{f}
 f = \left( \left.  \frac{1}{A} \, \frac{\d A}{\d x} \right|_{B_0}
\right)^{-1}
     \sum_{m=0}^\infty
     \left( \begin{array}{c} 1/n-1 \\ m \end{array} \right)
     \frac{(B-B_0)^m}{(B_0+C)^{m+1}}   \,.
\ee

After expanding the binomial $(B-B_0)^m$, we can solve the integral in
Eq.~(\ref{app}) for each of the terms of (\ref{f}) separately.
%
%
%
%
%
%
We obtain as our final result for general $B_0$ and $C>0$,
\be  \label{P_general}
 \ln\PLZ =  -\frac{\pi\gamma_n(B_0)}{2} \:  \F_n(\t,B_0)
\ee
with
\be  \label{gammaB}
 \gamma_n(B_0) =   \frac{\Delta S^2}{2E C \:|\d\ln A/\d x|_{B_0}}
\ee
and
\ba
 \F_n(\t,B_0) & = & \frac{2C}{B_0+C}\sum_{m=0}^\infty
 \left( \begin{array}{c} 1/n-1 \\ m \end{array} \right)
 \left(\frac{1}{B_0+C}\right)^m
\nonumber\\ & &
 \sum_{k=0}^m
 \left( \begin{array}{c} m \\ 2k \end{array} \right)
 \left( \begin{array}{c} 1/2 \\ k+1 \end{array} \right)
 (-B_0)^{m-2k} S^{2k} \,.
\ea
It is now clear that the choice of the point $x_0$ or $B_0$ where the
adiabaticity parameter $\gamma_n$ should be evaluated is totally
arbitrary.
A change of $B_0$ will always be compensated by a change in the
correction function $\F_n$ so that the physical observable $\PLZ$ is
independent of $B_0$, as it should.  The choice of the resonance point
$B_0=0$ is privileged only by the fact that it results in the simplest
analytical expression for $\F_n$ in the resonant region.
Indeed, by requiring the invariance of $\PLZ$ against variations of
$B_0$, one can write a differential equation for $\d\F_n/\d
B_0$~\cite{l}.
The solution of such equation allows to rescale the
correction functions obtained for $B_0=0$ to arbitrary $B_0$,
\be
 \F_n(\t,B_0)= \left(\frac{C}{B_0+C}\right)^{1/n}\F_n(\t,0) \,.
\ee
For practical calculations however it is simpler to insert $B_0=0$
whenever possible (resonance case) into Eq.~(\ref{P_general}) thereby
obtaining the well-known result of Ref.~\cite{Ku89},
\be
 \F_n(\t, 0)  = 2 \sum_{m=0}^\infty
 \left( \begin{array}{c} 1/n-1 \\ 2m \end{array} \right)
 \left( \begin{array}{c} 1/2 \\ m+1 \end{array} \right)
 \left( \tan(2\t) \right)^{2m} \,,
\ee
valid only for $\t<\pi/8$. Representing this series as a hypergeometric
function,
\be
 \F_n(\t, 0) = \,
 _2F_1\left( \frac{n-1}{2n}, \frac{2n-1}{2n};2;-\tan^2(2\t) \right) \,,
\ee
we can use the Euler integral representation~\cite{Pr90} of $_2F_1$,
\be
_2F_1(a,b;c;z) = \frac{\Gamma(c)}{\Gamma(b)\Gamma(c-b)} \;
   \int_0^1 \d t \: t^{b-1} (1-t)^{c-b-1} (1-tz)^a  \,,
\ee
as the analytical continuation for all $\t\in[0\!:\!\pi/4]$.

\subsection{The correction functions $\G_n$}
\label{sec:corr-funct-g_n}

In order to avoid the limited regime of validity of the formalism
presented in the previous subsection we will now present a new
representation for the crossing probability which is valid for all
$\t$.  We start directly from Eq.~(\ref{start}), but use now as
integration path in the complex $x$ plane the part of a circle of
radius $\Delta$ centered at zero and starting from $A_1=\Delta$ and
ending at
$A_2=\Delta\e^{2{\i}\t}$ namely, the end of the branch cut, see Fig.
\ref{path}.  In the case of a power-law potential $A=A_0 (r/R_0)^n$ by
substituting $x=R_0(\Delta/A_0)^{1/n} \e^{{\i}\varphi}$ we can factor
out the $\t$
dependence of $\PLZ$ into functions $\G_n$,
\be
 \ln\PLZ = -\kappa_n \G_n(\t) \,,
\ee
where
\be
 \kappa_n =
                 \left( \frac{\Delta}{E} \right) \:
                 \left( \frac{\Delta}{A_0} \right)^{1/n} R_0
\ee
is independent of $\t$ and
\be
 \G_n(\t) = \left| \; \Re
 \int_0^{2\t/n} \d\varphi \:\e^{{\i}\varphi}
 \left[\left(\e^{{\i} n\varphi}-C\right)^2 + S^2 \right]^{1/2}  \right|
\,.
\ee
The functions $\G_n$ are well suited for numerical evaluation and
always correspond to a neutrino state propagating in the physical part
of the $x$ plane, $x>0$. Therefore, they have, in contrast to the
$\F_n$ functions, the period $\pi/2$ and are valid for all $\t$.
Anti-neutrinos feel a potential $A$ with the
opposite sign than neutrinos. This sign change can be compensated by
the exchange of $\t$ with $\pi/2-\t$; thus the formulae derived for
the crossing probability of neutrinos become valid for anti-neutrinos
substituting $\t\to\pi/2-\t$.

Finally, we remind the reader that the Landau-St{\"u}ckelberg-Zener
approach is only valid for small deviations from adiabaticity.
Therefore, in order to cover also the non--adiabatic case, the WKB
formula Eq.~(\ref{WKB0}) has to be replaced~\cite{Ku89} by
\be    \label{xxx}
 P_c = \frac{\exp\left(-\kappa_n\G_n\right) -
             \exp\left(-\kappa_n\G_n^\prime\right)}
            {1 - \exp\left(-\kappa_n\G_n^\prime\right)} \,,
\ee
where $\G_n^\prime=\G_n/\sin^2\t$ for neutrinos and
$\G_n^\prime=\G_n/\cos^2\t$
for anti-neutrinos, respectively.  A similar formula holds for the
$\F_n$ functions, if $\kappa_n\G_n$ is replaced by $\gamma\F_n\pi/2$.

\section{Neutrino oscillations in supernova envelopes}
\label{sec:neutr-oscill-supern}

\subsection{General discussion}
\label{sec:general-discussion}

The potential profile $A(r)$ in supernova (SN) envelopes can be
approximated by a power law with $n\ap -3$, and
$V(r)= 1.5\times 10^{-9}$~eV $(10^9~{\rm cm}/r)^{3}$~\cite{SN}.
Since only $\bar\nu_e$  were detected from SN~1987A and also in the
case of a future galactic SN the $\bar\nu_e$ flux will dominate the
observed neutrino signal, an analytical expression  for $P_c$ valid in
the
non-resonant part of the mixing  space is especially useful. In the
following, we will always consider oscillations of anti-neutrinos.

The probability for a $\bar\nu_e$ to arrive at the surface of the Earth
can be written as an incoherent sum of probabilities,
\be
\label{Pee}
  P_{\bar e \bar e} =
  P^S_{\bar e1} P^E_{1\bar e} + P^S_{\bar e2} P^E_{2\bar e} =
  \left(1-P_c\right)\cos^2\t + P_c\sin^2\t \,,
\ee
where $P^S_{\bar ei}$ denotes the probability that a $\bar\nu_e$ leaves
the
star as mass eigenstate $\bar\nu_i$ and $P^E_{i\bar e}$ the probability
that $\bar\nu_i$ is detected as $\bar\nu_e$.

In Fig.~\ref{SN}, we compare the results of a numerical solution of
the Schr{\"o}dinger equation~(\ref{S}) with the analytical calculation
of
$P_{\bar e \bar e}$ using the $\G_{-3}$ and the $\F_{-3}$ functions.
The latter is shown only for its range of applicability, $\t>\pi/4$.
The agreement between the two methods using the WKB approach is again
(for $\t>\pi/4$) excellent. Generally, these two methods agree also very
well with the results of the numerical solution of the Schr{\"o}dinger
equation; there are only tiny deviations in the region $\Delta/(2E)\sim
10^{-17}$~eV.

Next, we discuss the quality of the different approximations for
$\PLZ$ which have so far been used in the literature. The most common
one is to use the correct potential profile, $A\propto r^{-3}$, together
with $\F_1=1$~\cite{snbounds}. Since all correction functions $\F_n$
go to 1 for small mixing in the resonant region, this approximation is
certainly justified for $\t\to 0$ in neutrino and for $\t\to\pi/2$ in
anti-neutrino oscillations.  The shape of the MSW triangle which is
determined\footnote{Note that, according to Eq.~(\ref{gamma0}), the
  hypotenuse of the MSW triangle should have the slope $n/(n+1)$ in a
  $\ln(\Delta)$--$\ln(\tan^2\t)$ plot.} by the exponent $n$ is correctly
reproduced in this approximation, as the comparison of the numerical
results and the linear approximation in Fig.~\ref{linear} confirms.
However, this approximation becomes worse for $\tan^2\t\lsim 5$ and
breaks down for $\tan^2\t<1$. Therefore, it is not applicable in the
phenomenologically most interesting region of maximal or nearly
maximal mixing presently indicated by the solar neutrino
data~\cite{solar,solar2,Gonzalez-Garcia:2001sq}.
Physically, the failure of the linear approximation \cite{snbounds}
already in the resonant region is understandable by the broadening of
the crossing probability profile, cf.~Fig.~\ref{pmva2}.  Since for
larger mixing the crossing probability is determined not only by the
potential near the expansion point but by a rather large, extended
region, a linear approximation of the potential profile is not good
enough.

In order to obtain an approach well-defined for all $\t$ another
approximation was employed in Ref.~\cite{Ka01}. There, the expression
$\gamma_n\F_n$ valid for an exponential profile was used. The resulting
shape of the MSW triangle outside the region of nearly-maximal mixing
is not correctly reproduced because of the wrong profile used, as can
be seen from Fig.~\ref{f_exp2}. The agreement between $P_c$ from the
numerical solution of the Schr{\"o}dinger equation and that obtained by
using the exponential profile is however very good for $\tan^2\t\lsim
5$ and a suitable choice of $A_0$ and $R_0$. Therefore, this
approximation is adequate for the discussion of {\em anti-}neutrino
oscillations in the whole parameter space allowed by present solar
neutrino experiments. 
Numerically, we have found the best agreement between the two
approaches for $R_0^{\rm exp}\ap 100\times R_0^{\rm power}$, where we
have
fixed $A_0^{\rm exp}=A_0^{\rm power}$.  This normalization of the
scale factor $R_0$ ensures that both potentials,
$A=A_0\exp(-r/R_0^{\rm exp})$ and $A=A_0(r/R_0^{\rm power})^{-3}$,
give rise to the same scale height, $[\d\ln A(r)/\d r]^{-1}$, in the
region around the PMVA for the values of $\t$ of interest.
Figure~\ref{f_exp2} shows very good agreement between this
approximation and the numerical results except for $\tan^2\t\gsim 3$
and $\Delta/E\gsim 10^{-10}$~eV$^2$/MeV.

Reference~\cite{Ku89} gave an additional approximation for $\F_{-3}$
which consists of an expansion of the exponential profile in the
parameter $1/n$,
\ba \label{KP}
 \F_n(\t) &=& (1-\tan^2\t)\: \big\{ 1-\frac{1}{n} \big[
 \ln (1-\tan^2\t) + 1
\nonumber\\ & &
 - \frac{(1+\tan^2\t)}{\tan^2\t} \ln(1+\tan^2\t) \big] +\ldots \big\}
\,.
\ea
This approximation agrees indeed very well already for $n=-3$ with our
numerical results in the resonant region, cf.~Fig.~\ref{linear}. Only
near the limit of its range of applicability, $\tan^2\t =1$, sizeable
deviations can be seen.

Finally, we have also checked how deviations of the true SN progenitor
profile $V(r)$ from an $1/r^3$ profile may affect our results for
$P_{\bar e\bar e}$. Realistic progenitor profiles differ in two
aspects from the simple $1/r^3$ profile assumed. First, the outer part
of the envelope has an onion like structure, and its chemical
composition, $Y_e(r)$, and thus also $V(r)$ changes rather sharply at
the boundaries of the various shells. Second, the density drops faster
in the outermost part of the envelope, becoming closer to an
exponential decrease.  We have calculated numerically $P_{\bar e\bar
  e}$ using profiles for three different progenitor masses (11, 20 and
30~$M_\odot$) from Woosely~\cite{Wo} and one by Nomoto~\cite{No} for the
progenitor of SN~1987A.  We have found that $P_{\bar e\bar e}$ is well
approximated in the non-resonant part by our analytical results for
the $1/r^3$ profile, independently of the details of the progenitor
profile. In contrast, $P_{\bar e\bar e}$ depends strongly on the
details of the progenitor profile in the resonant region. In
particular, there are rather strong deviations between a $1/r^3$
profile and the profiles of Refs.~\cite{Wo,No}. As an example, we
compare in Fig.~\ref{woosley} the $P_{\bar e\bar e}$ calculated
numerically for the $20M_\odot$ profile of Ref.~\cite{Wo} with the
analytical results for our standard SN profile $V(r)= 1.5\times
10^{-9}$~eV
$(10^9~{\rm cm}/r)^{3}$.  We conclude that a numerical solution of the
Schr{\"o}dinger equation~(\ref{S}) should be performed in the resonant
region, using a realistic profile for the particular progenitor star
considered.  However, a $1/r^3$ profile together with the LSZ crossing
probability is sufficient for the analysis of {\em anti-}neutrino
oscillations in the phenomenologically most interesting region
$\tan^2\t\lsim 5$. Note that it is this region where the other methods
based on the splitting of $\PLZ$ into the adiabaticity factor $\gamma$
plus the functions $\F$ fail.

\subsection{Likelihood analysis of the  SN~1987A neutrino signal}
\label{sec:likel-analys-sn}

In Ref.~\cite{Ka01} we performed a likelihood 
analysis~\cite{Jegerlehner:1996kx} of the neutrino signal of SN~1987A
observed by the Kamiokande~II and the IMB experiments~\cite{obs} using
the expression $\gamma_n\F_n$ valid for an exponential potential profile
for $\PLZ$.  In order to describe the SN~1987A neutrino signal we need
the evolution of a typical SN anti-neutrino in the region of
parameters presently indicated by the solar neutrino
problem~\cite{solar,solar2}. Most of the presently allowed oscillation
solutions of the solar neutrino problem (LMA--MSW, LOW-QVAC, VAC,
``just-so'') require large neutrino mixing.  The small mixing angle
solution, called SMA-MSW, appears in most analyses only at about 99.9\%
C.L. The precise value slightly depends on arbitrary details of the
statistical analysis, as can be seen by comparing different global
analyses of the solar data.  So far, only the SuperKamiokande
collaboration can perform a full analysis of its signal and
background~\cite{hep-ex/0103033}.  However, the published data allow
to approximate reasonably well the full SuperKamiokande results so as
to be used in global analyses.  We have here adopted the values of
the $\chi^2_\odot(\vartheta,\Delta)$ grid corresponding to Fig.~6b of
Ref.~\cite{solar}.

The description of the anti-neutrino conversion is trivial both in the
cases of complete adiabaticity or full non-adiabaticity. In the first
case there is, of course no-level crossing, while in the second the
standard vacuum treatment applies. In either case the exact form of
the potential profile used has no effect.  As we have seen in
Figs.~\ref{border1} and \ref{border2} full adiabaticity or
non-adiabaticity occur for an exponential and a power-law like profile
with $n=-3$ for all solutions to the solar neutrino problem, except
for the case of LOW-QVAC solution.
Therefore for the former solutions the main part of the results and
conclusions of Ref.~\cite{Ka01} remains unchanged.  However, for the
LOW-QVAC a new analysis is in order.

Here we have repeated therefore the likelihood fit of the neutrino
signal from SN~1987A in the $\tan^2\t$--$\Delta$ plane for a more
realistic $1/r^3$ profile together with the expression (\ref{xxx}) for
the crossing probability but used otherwise exactly the same approach 
as in Ref.~\cite{Ka01}. In particular, we work in the framework of
oscillations between two {\em active\/} neutrino flavours.
Our 2-flavour approximation is valid for the case of normal mass
hierarchy
$m_3  \gg  m_1\sim m_2$ in view of the Chooz limit~\cite{chooz} on
$U_{13}$. In the alternative inverse mass hierarchy case $m_1\sim m_2
\gg m_3$, the third
neutrino do not significantly affect supernova oscillations
if $s_{13}^2< \hbox{few}\times 10^{-4}$ because then the ``heavy''
resonance is non-adiabatic.
With the inverted spectrum, larger values of $s_{13}$ are already
strongly
disfavoured by SN~1987A~\cite{Minakata:2001rx}.

The ratio of the likelihood functions ${\cal L}$ for different hypotheses
measures the degree to which the experimental data favour one
hypothesis over the other. In order to decide how strong the large
mixing solutions are disfavoured with respect to the no-oscillation case
(or to SMA--MSW oscillations, that negligibly affect supernova $\bar\nu_e$)
for a certain choice of astrophysical parameters $\{\alpha\}$,
we consider the ratio
\be  \label{R}
 R_\alpha(\t,\Delta) = \frac{{\cal L}_\alpha(\t,\Delta)}
                        {{\cal L}_{\alpha,{\rm NO-OSC}}}
\ee
as function of $\t$ and $\Delta$. The likelihood function 
\be
 {\cal L}_\alpha(\t,\Delta) \propto \exp\left(-\int n_\alpha(E) \; dE \right) 
               \prod_{i=1}^{N_{\rm obs}} n_\alpha(E_i,\t,\Delta)
\ee
tests the hypothesis that a prescribed neutrino fluence
$F_{\bar\nu_e}$ leads to the observed
experimental data $E_i$ with probability distribution 
$n_\alpha(E,\t,\Delta)$.
The connection between the observed data set $E_i$ and the emitted
neutrino fluence $F$ has been
described already in detail in the literature, cf. Ref.~\cite{Ka01}.

The astrophysical parameters $\alpha$ used
are the released binding energy $\Eb$, the average energy $\Ee$ of
$\bar\nu_e$ neutrinos, and the average energy $\Eh=\tau\Ee$ of
$\bar\nu_{\mu,\tau}$ neutrinos.  For a more details see
Ref.~\cite{Ka01}.  We consider mainly two sets of values for
$\{\Eb,\Ee\}$: one corresponds to the 
lowest values found in simulations, $\Eb=1.5\times 10^{53}$~erg and
$\Ee=12$~MeV, and the second one is with $\Eb=3\times 10^{53}$~erg and
$\Ee=14$~MeV nearer to average values found~\cite{simu}.

In Figs.~\ref{1.5_12_sep} and \ref{3_14_sep} we show the likelihood
ratio $\ln(R)$ as function of $\tan^2\t$ and $\Delta$ relative to the
NO-OSC hypothesis. The contours of constant likelihood shown
correspond to $\ln(R)=-1,-2,-3,-5,-10,-15,-20$, unless otherwise
indicated. There are two differences compared to the same plots of
Ref.~\cite{Ka01}. First, since in Ref.~\cite{Ka01} the scale height
$R_0$ of the exponential profile was not optimized to reproduce best a
power-law profile with $n=-3$, matter effects in the SN envelope
became important at $10^{-7}$-$10^{-8}$~eV$^2$ while here we find that
they already start being important for ${\rm few} \times
10^{-9}$~eV$^2$.
Second, the slope of the MSW triangle in the dark side $\tan^2\t>1$
has now the correct value.  The changes in the regions favoured by the
solar neutrino data are however marginal.

\subsection{Combined global analysis of solar and SN~1987A
  neutrino data}
\label{combinedanalysis}

In Figs.~\ref{1.5_12_sep} and \ref{3_14_sep}, we have simply
superimposed the C.L. contours obtained by analyzing solar and SN data
separately. Since the two data sets are statistically independent and
functions of the same two fit parameters, they can be trivially
combined,
\be
 \chi^2_{\rm TOT}(\t,\Delta) =
 \chi^2_\odot(\t,\Delta)+\chi^2_{\rm SN}(\t,\Delta) \,.
\ee
We can then obtain new C.L. contours which are defined relative to the
minimum of $\chi^2_{\rm TOT}$. In the following, we will use the
$\chi^2_\odot(\t,\Delta)$ grid calculated in Ref.~\cite{solar} for the
solar data and $\chi^2_{\rm SN}=-2\, {\cal L}(\t,\Delta)$ with
${\cal L}(\t,\Delta)$ as
defined in Ref.~\cite{Ka01} for the SN~1987A data.  We consider the
astrophysical parameters as known and minimize only the two parameters
$\t$ and $\Delta$. Hence, the confidence levels are always calculated
with
respect to two fit parameters.

\begin{center}
    \begin{figure}
\epsfig{file=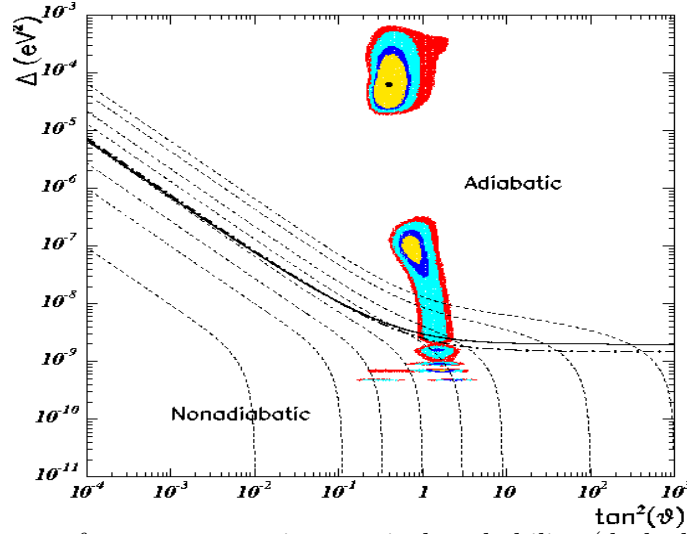,height=7.cm,width=9.cm,angle=0}
\caption{\label{border1}
  Contours of constant neutrino survival probability (dashed) together
  with the borderlines Eq.~(\ref{crit}) (solid) and (\ref{fried})
  (dash-dotted) between adiabatic and non-adiabatic regions for the
  solar density profile.  The 90, 95, 99 and 99.73\% C.L. contours for
  the different solutions to the solar neutrino problem are also
  shown.}
\end{figure}
  \end{center}

\begin{center}
\begin{figure}
\epsfig{file=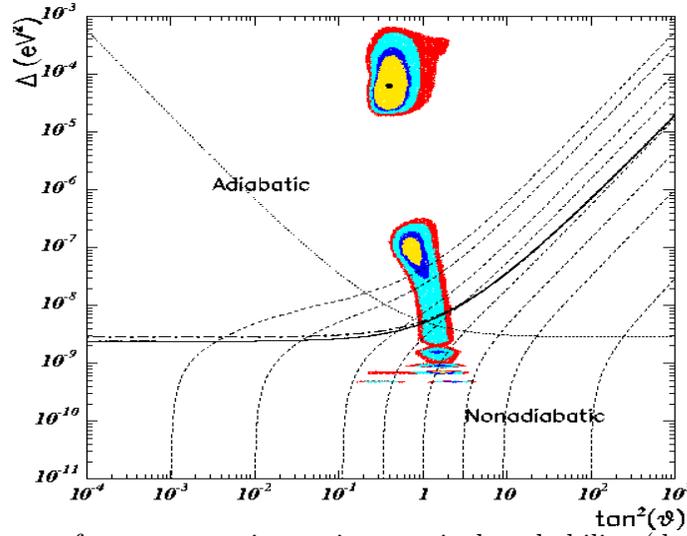,height=7.cm,width=9.cm,angle=0}
\caption{\label{border2}
  Contours of constant anti-neutrino survival probability (dashed)
  together with the borderlines Eq.~(\ref{b}) between adiabatic and
  non-adiabatic regions using Eq.~(\ref{i}) (solid) and (\ref{ft})
  (dash-dotted) for $f(\t)$ for the SN profile given in the text; the
  dotted line shows the borderline for neutrinos.  The 90, 95, 99 and
  99.73\% C.L. contours for the different solutions to the solar
  neutrino problem are also shown.}
\end{figure}
\end{center}

\begin{center}
\begin{figure}
\epsfig{file=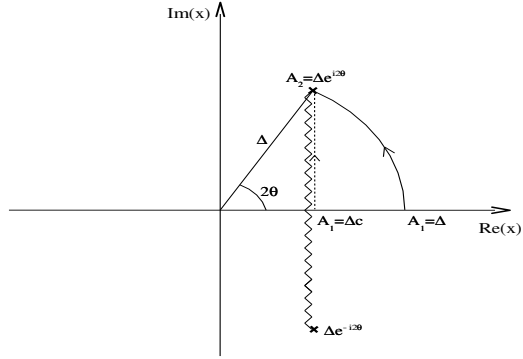,height=7.cm,width=9.cm,angle=0}
\caption{\label{path}
Analytical structure of the function $\Dm$ together with the different
integration paths used in Sec.~\ref{sec:corr-funct-g_n} for the
special case $n=1$. The branch cut is shown by a wavy line, poles by
crosses; an additional pole appears for $n<0$ at $x=0$.}
\end{figure}
\end{center}

\begin{center}
\begin{figure}
  \epsfig{file=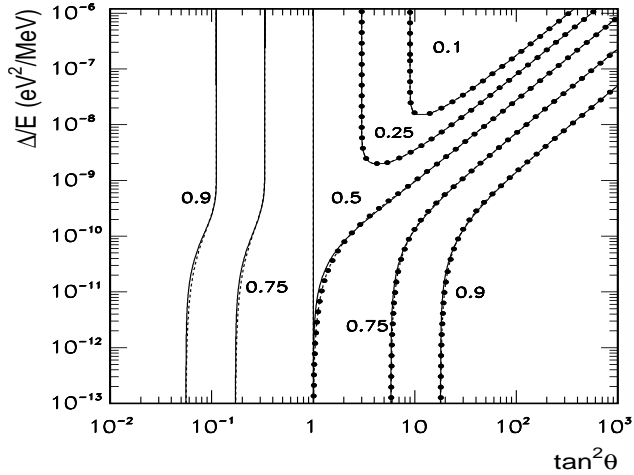,height=7.cm,width=9.cm,angle=0}
\caption{\label{SN}
  Contours of constant survival probability $P_{\bar e\bar e}$,
  numerically (solid lines), with $\G_{-3}$ (dashed lines) and
  $\F_{-3}$ (filled circles, only for $\t>\pi/4$), for $A\propto
  r^{-3}$ as given in the text.}
\end{figure}
\end{center}

\begin{center}
\begin{figure}
\epsfig{file=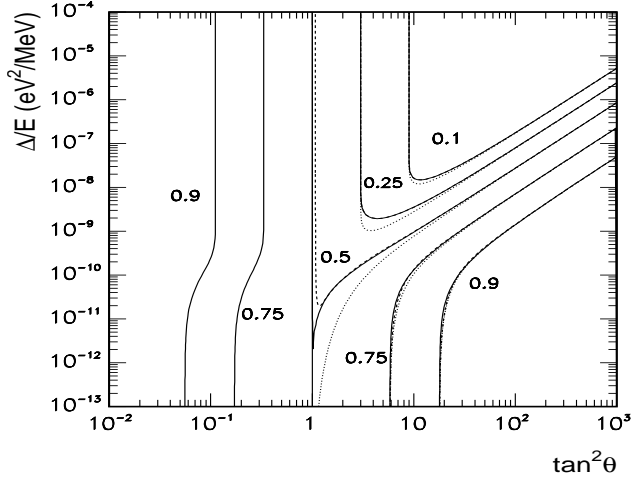,height=7.cm,width=9.cm,angle=0}
\caption{\label{linear}
  Comparison of the contours of constant survival probability $P_{\bar
  e\bar e}$ calculated numerically (solid lines) and with the linear
  approximation $\F_{1}=1$ (dashed lines) and the approximation
  Eq.~(\ref{linear}) (dotted lines), respectively; 
  all for $A\propto r^{-3}$ as given in the text.}
\end{figure}
\end{center}

\begin{center}
\begin{figure}
\epsfig{file=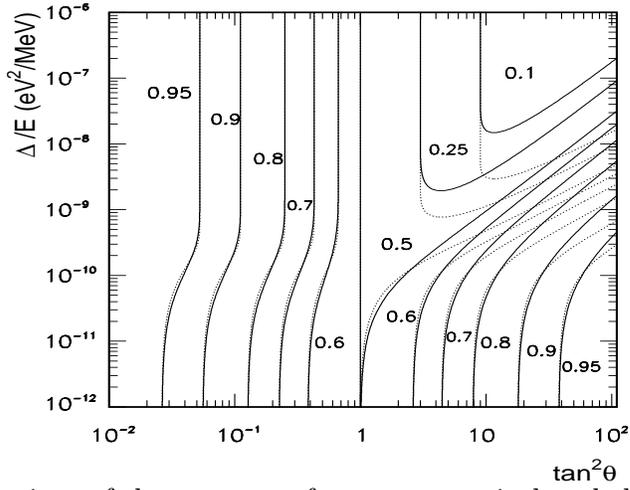,height=7.cm,width=9.cm,angle=0}
\caption{\label{f_exp2}
Comparison of the contours of constant survival probability
$P_{\bar e\bar e}$ calculated numerically (solid lines)
for $A\propto r^{-3}$
and with the exponential approximation $A\propto\exp(-r/R_0^{\rm exp})$
(dashed lines) with $R_0^{\rm exp} = 100\times R_0^{\rm power}$.}
\end{figure}
\end{center}

\begin{center}
\begin{figure}
\epsfig{file=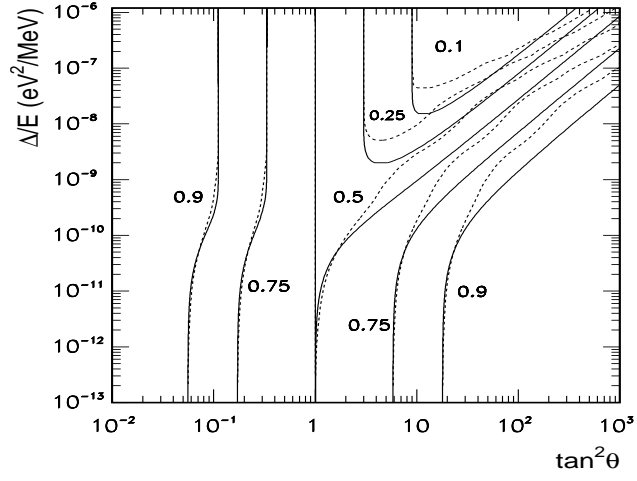,height=7.cm,width=9.cm,angle=0}
\caption{\label{woosley}
Comparsion of the contours of constant survival probability $P_{\bar
    e\bar e}$ calculated numerically for a $M=20M_\odot$ progenitor star
from Ref.~[17] (dotted lines) and calculated for $A\propto r^{-3}$
  with the $\PLZ$ approximation (solid lines). }
\end{figure}
\end{center}

\begin{center}
\begin{figure}
\hspace*{-0.4cm}
\epsfig{file=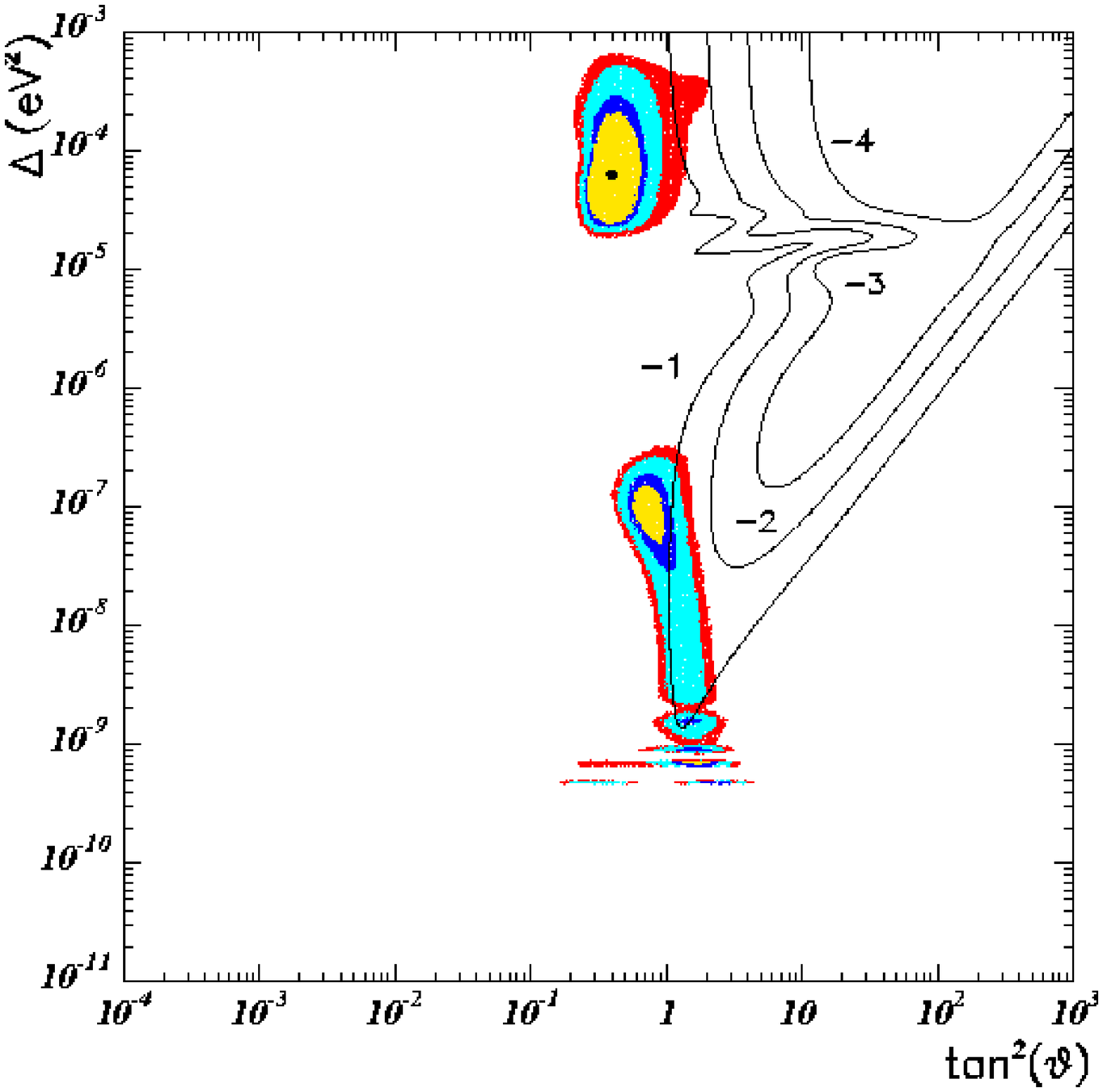,height=7.cm,width=9.cm,angle=0}
\epsfig{file=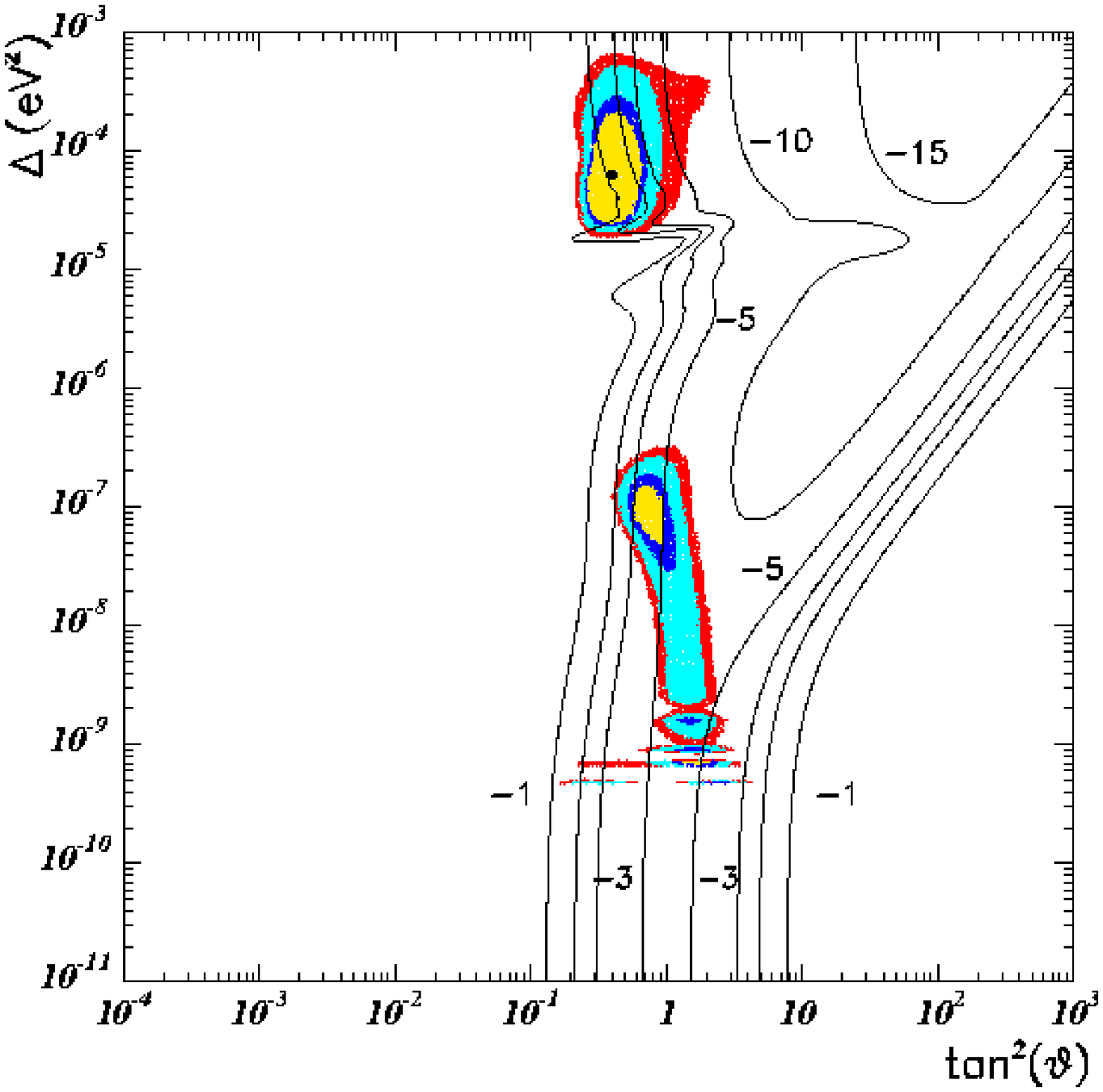,height=7.cm,width=9.cm,angle=0}
\epsfig{file=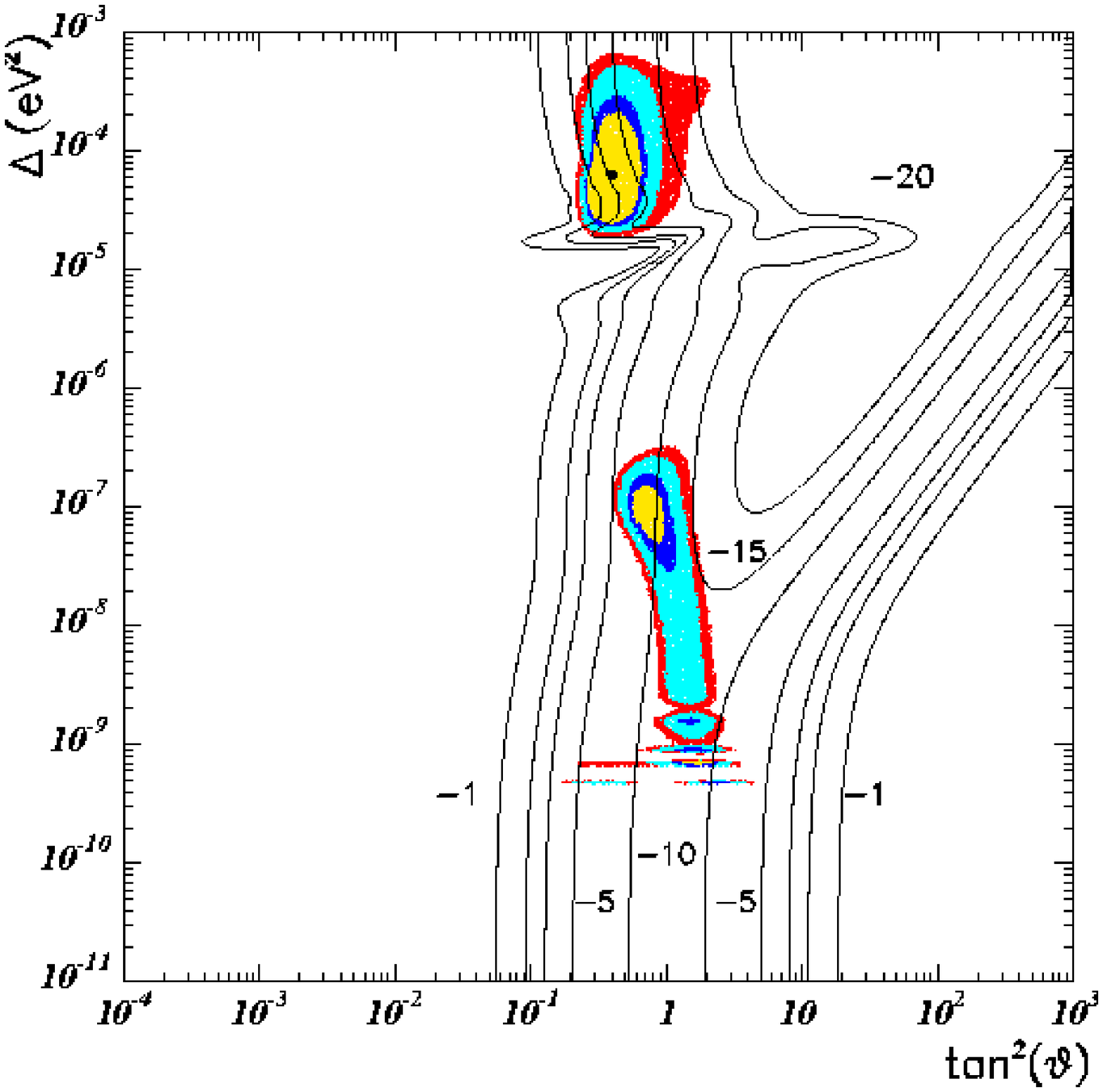,height=7.cm,width=9.cm,angle=0}
\caption{\label{1.5_12_sep}
  Likelihood ratio $\ln(R)$ relative to the NO--OSC
  hypothesis, as function of $\tan^2\t$ and $\Delta$/eV$^2$
  for $\tau=\Eh/\Ee=1.4$ (top), $\tau=1.7$ (middle) and $\tau=2$
  (bottom) together with
  the different solutions to the solar neutrino problem.
  All figures for $\Eb=1.5 \times 10^{53}$~erg and $\Ee=12$~MeV.}
\end{figure}
\end{center}

\begin{center}
\begin{figure}
\hspace*{-0.4cm}
\epsfig{file=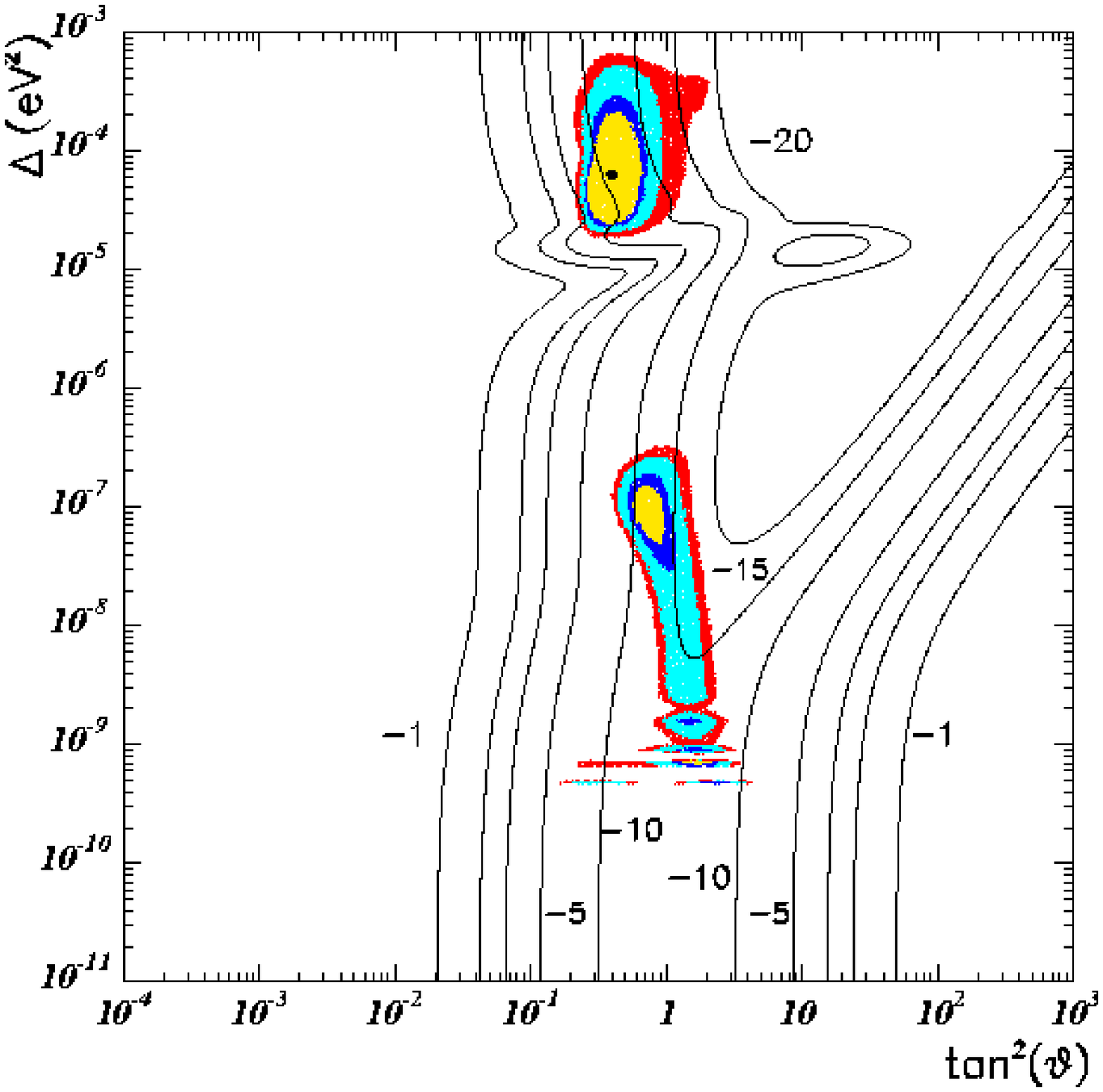,height=7.cm,width=9.cm,angle=0}
\epsfig{file=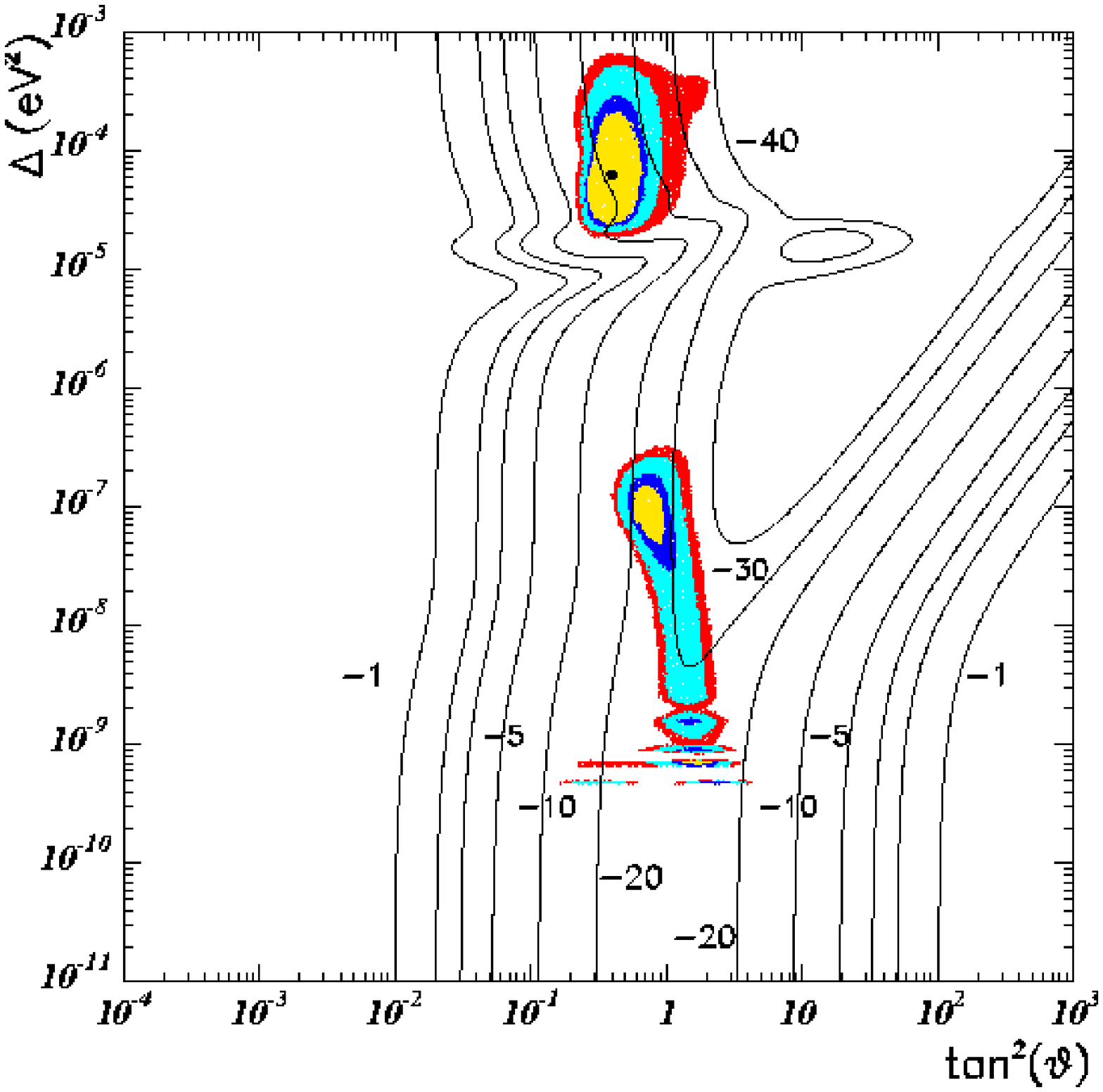,height=7.cm,width=9.cm,angle=0}
\epsfig{file=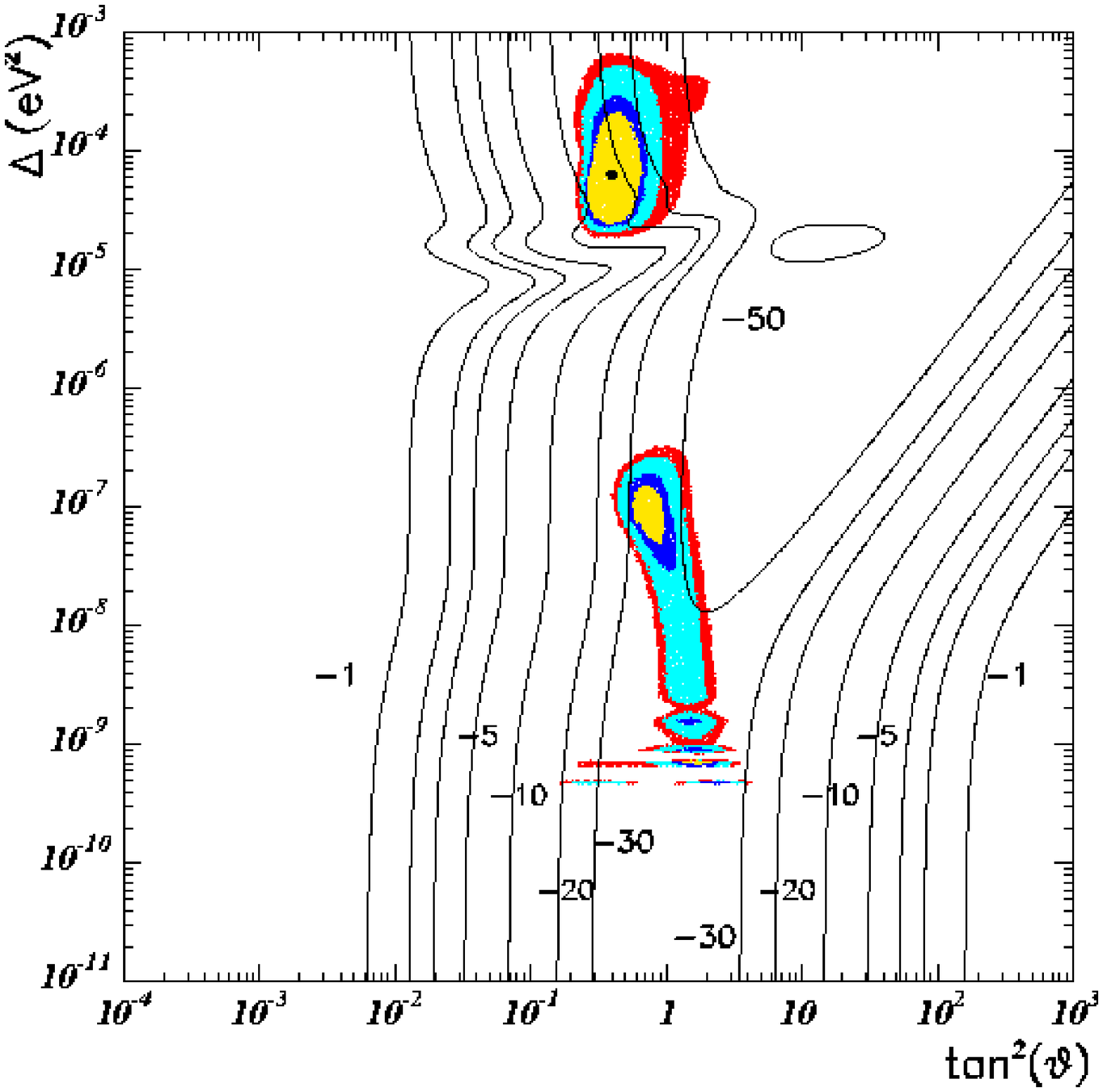,height=7.cm,width=9.cm,angle=0}
\caption{\label{3_14_sep}
  Likelihood ratio $\ln(R)$ relative to the NO--OSC
  hypothesis, as function of $\tan^2\t$ and $\Delta$/eV$^2$
  for $\tau=\Eh/\Ee=1.4$ (top), $\tau=1.7$ (middle) and $\tau=2$
  (bottom) together with
  the different solutions to the solar neutrino problem.
  All figures for $\Eb=3\times 10^{53}$~erg and $\Ee=14$~MeV.}
\end{figure}
\end{center}

\begin{center}     
\begin{figure} 
\hspace*{-0.4cm}
\epsfig{file=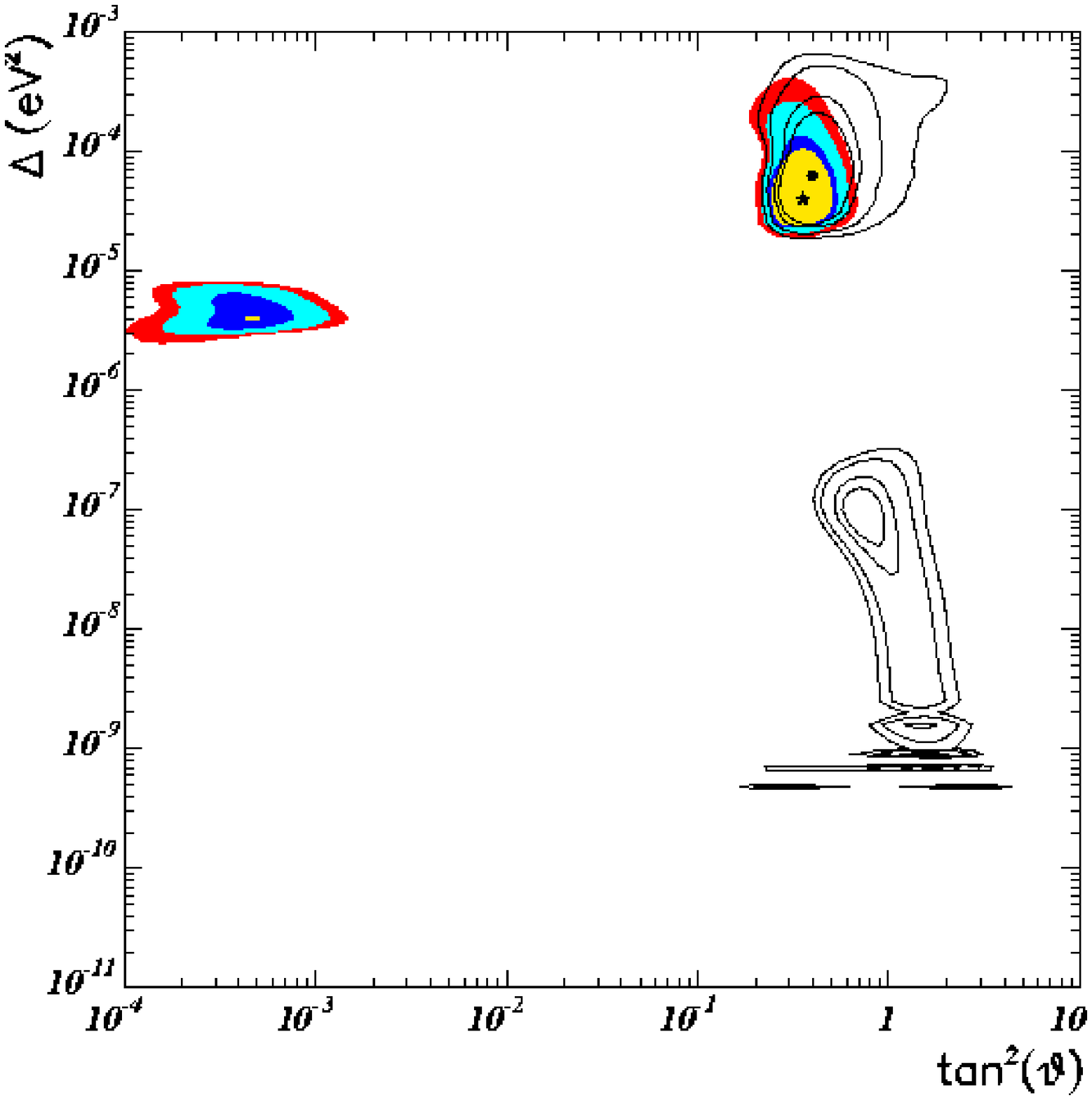,height=7.cm,width=9.cm,angle=0}
\epsfig{file=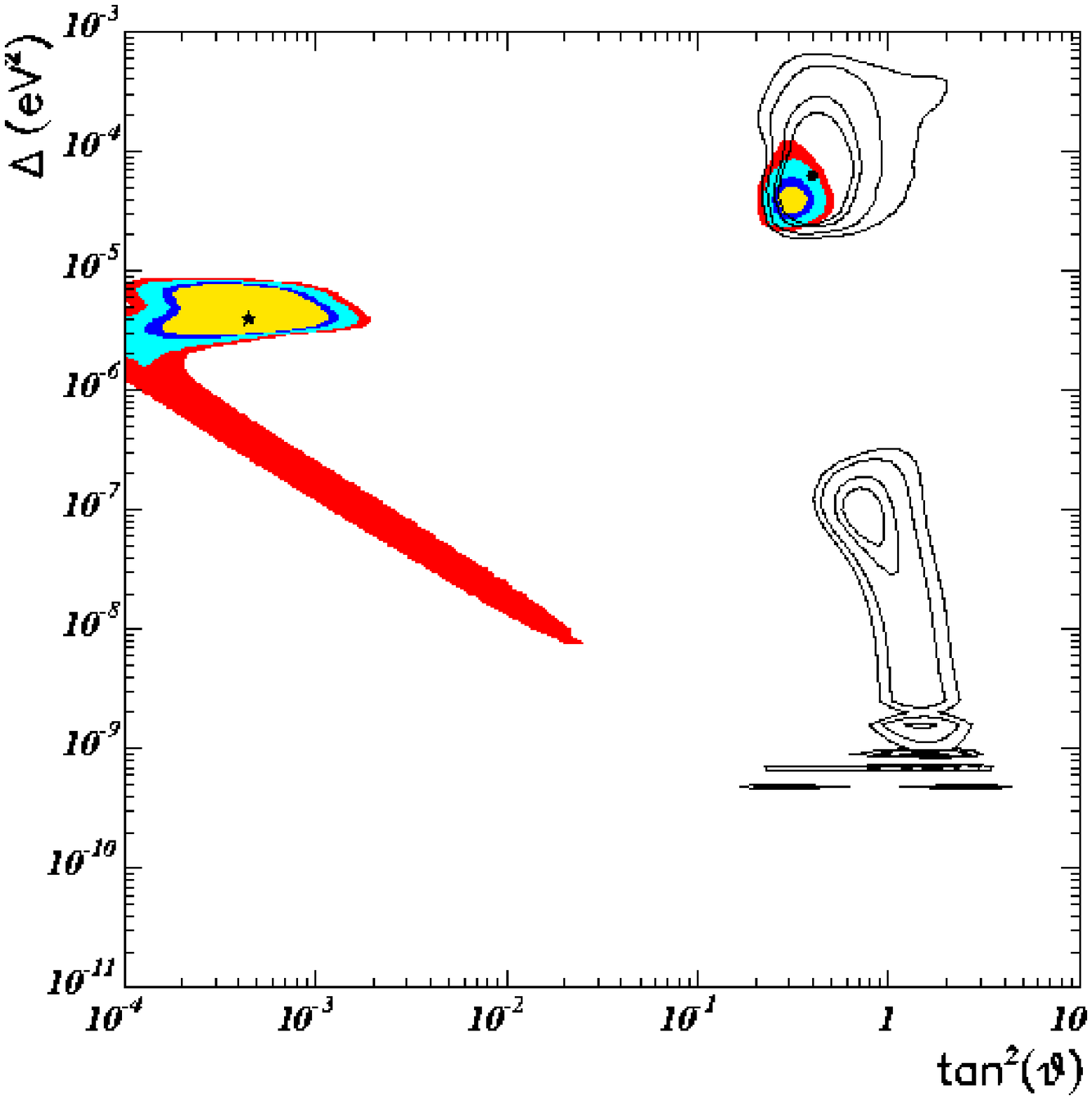,height=7.cm,width=9.cm,angle=0}
\epsfig{file=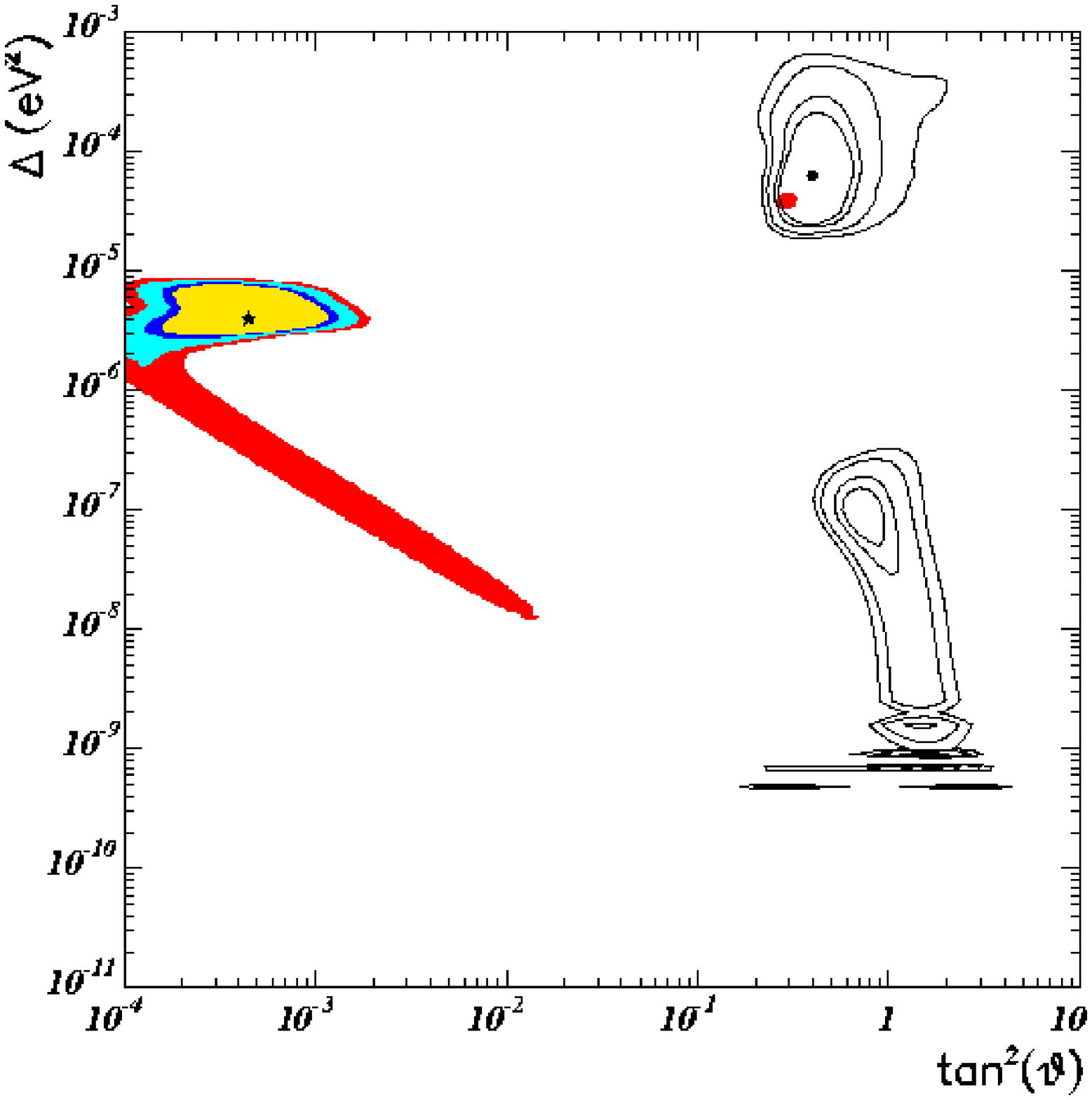,height=7.cm,width=9.cm,angle=0}
\caption{\label{3_14_tot}
  The 90, 95, 99 and 99.73\% C.L. contours of the combined fit of solar
and
  SN~1987A data (coloured/grey) together with the contours of the solar
  data alone (solid lines);
  for $\tau=\Eh/\Ee=1.4$ (top), $\tau=1.7$ (middle) and $\tau=2$
(bottom).
  All figures for $\Eb=3\times 10^{53}$~erg and $\Ee=14$~MeV.}
\end{figure}
\end{center}

\begin{center}     
\begin{figure}
\hspace*{-0.4cm}
\epsfig{file=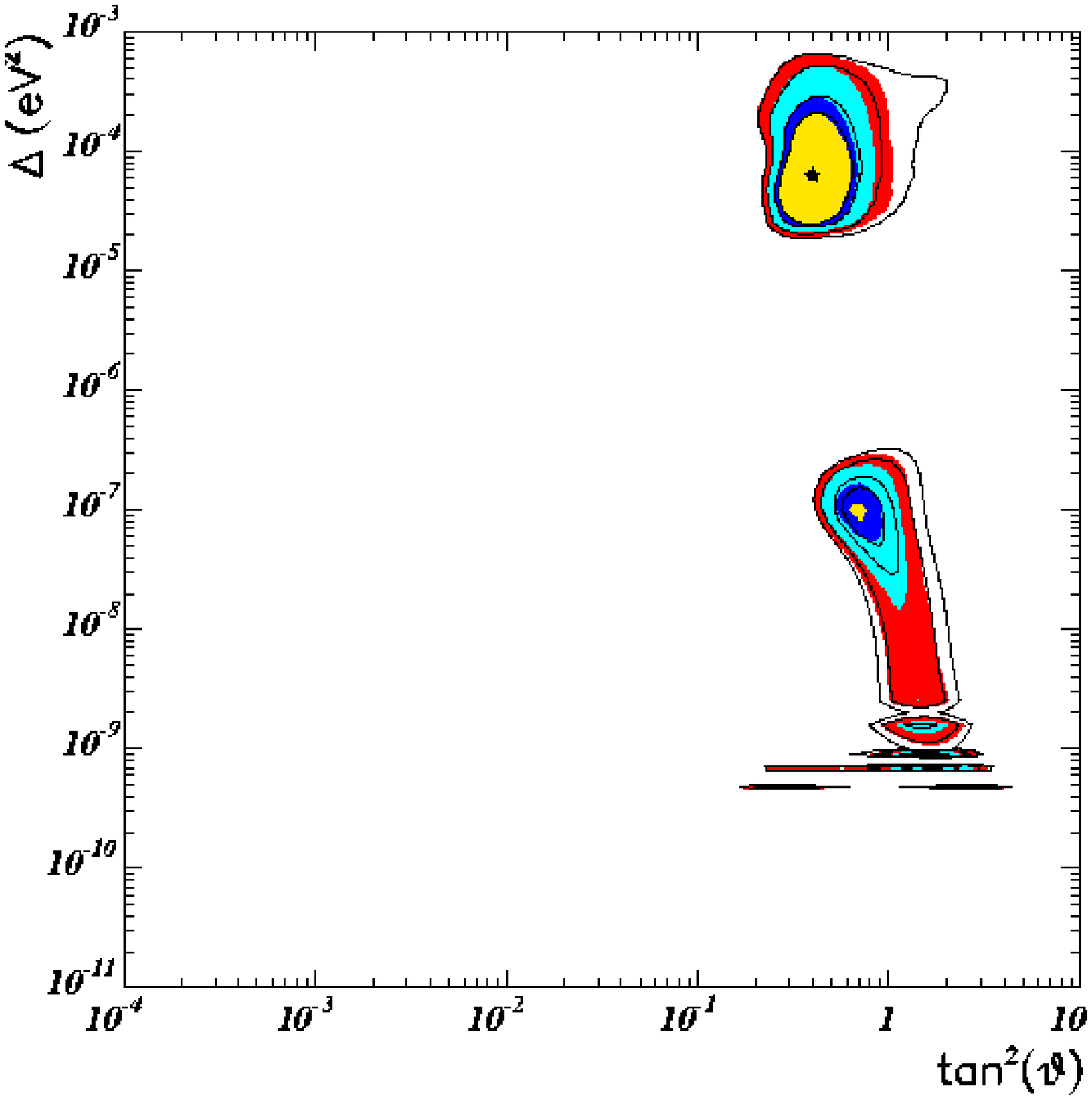,height=7.cm,width=9.cm,angle=0}
\epsfig{file=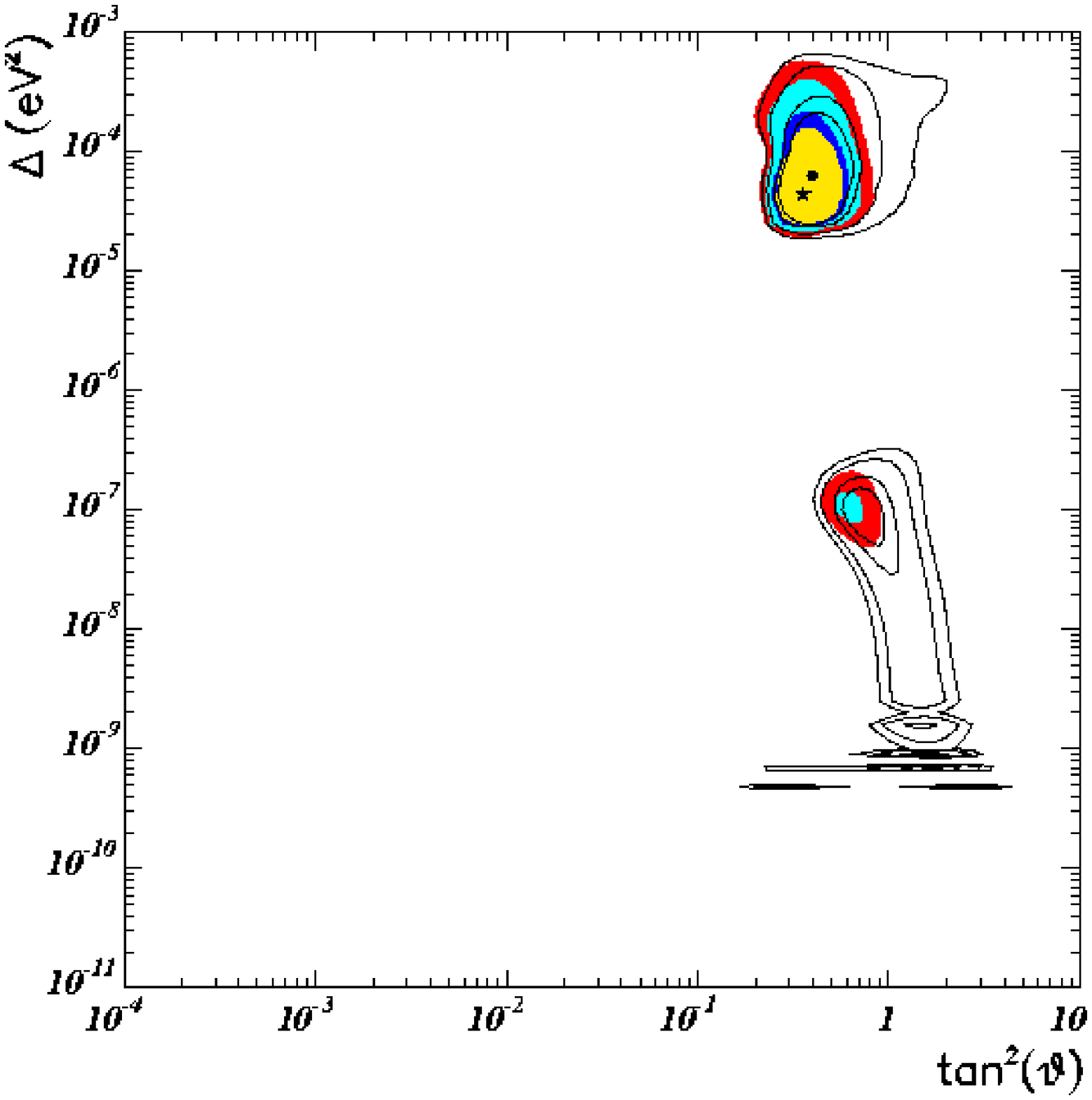,height=7.cm,width=9.cm,angle=0}
\epsfig{file=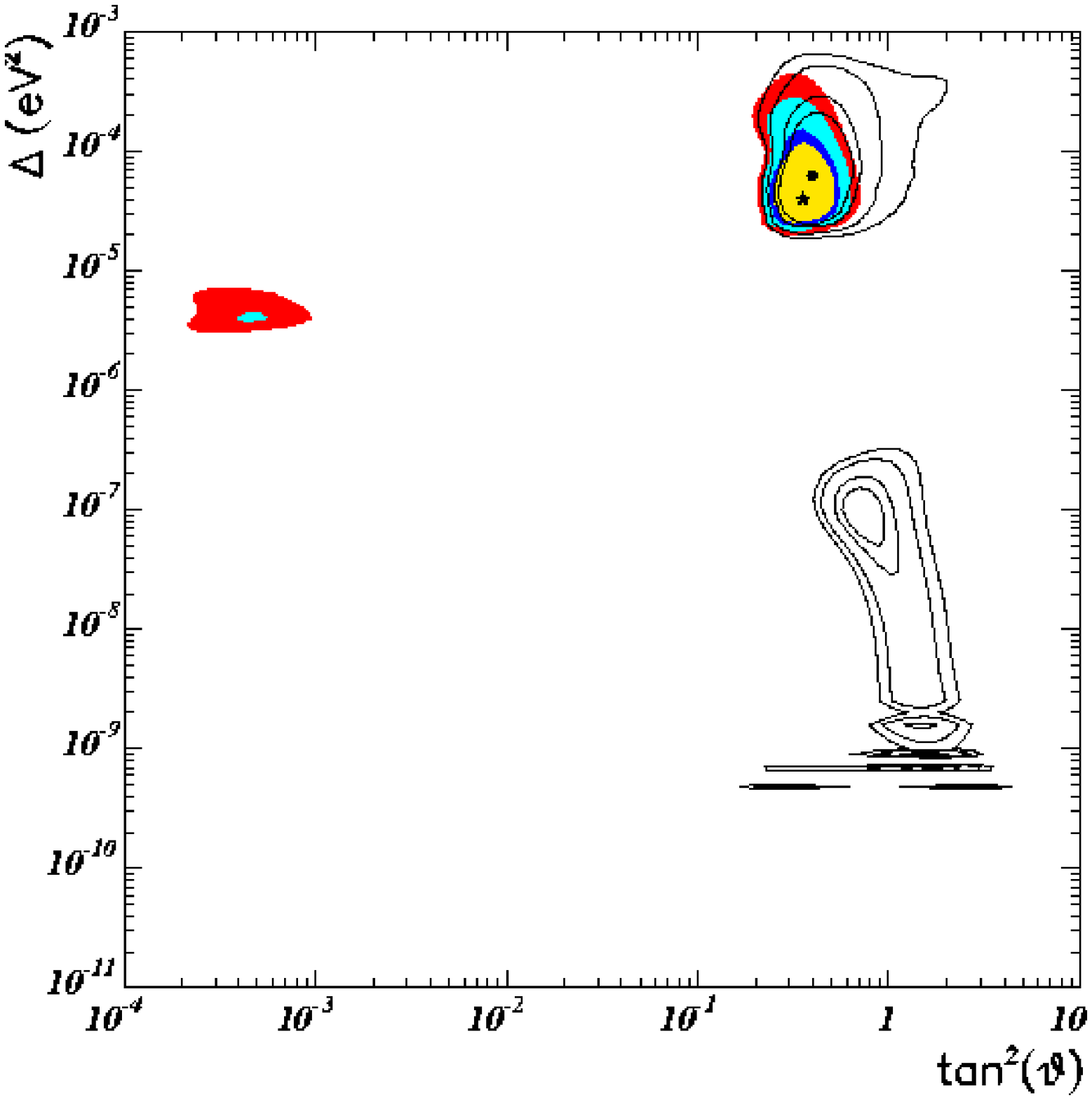,height=7.cm,width=9.cm,angle=0}
\caption{\label{1.5_12_tot}
  The 90, 95, 99 and 99.73\% C.L. contours of the combined fit of solar
and
  SN~1987A data (coloured/grey) together with the contours of the solar
  data alone (solid lines);
  for $\tau=\Eh/\Ee=1.4$ (top), $\tau=1.7$ (middle) and $\tau=2$
(bottom).
  All figures for $\Eb=1.5 \times 10^{53}$~erg and $\Ee=12$~MeV.}
\end{figure}
\end{center}

\begin{center}    
\begin{figure}
\hspace*{-0.4cm}
\epsfig{file=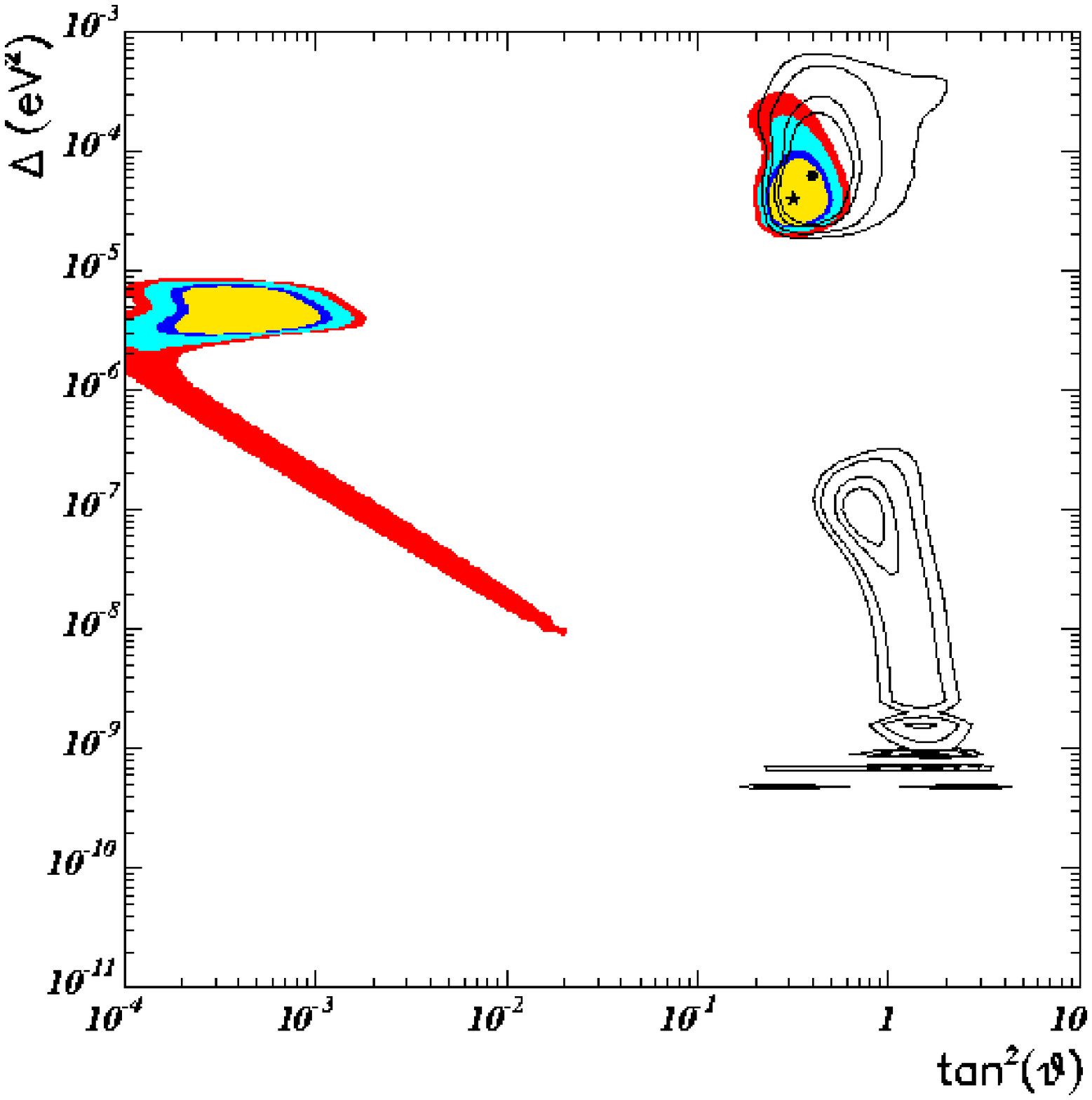,height=7.cm,width=9.cm,angle=0}
\epsfig{file=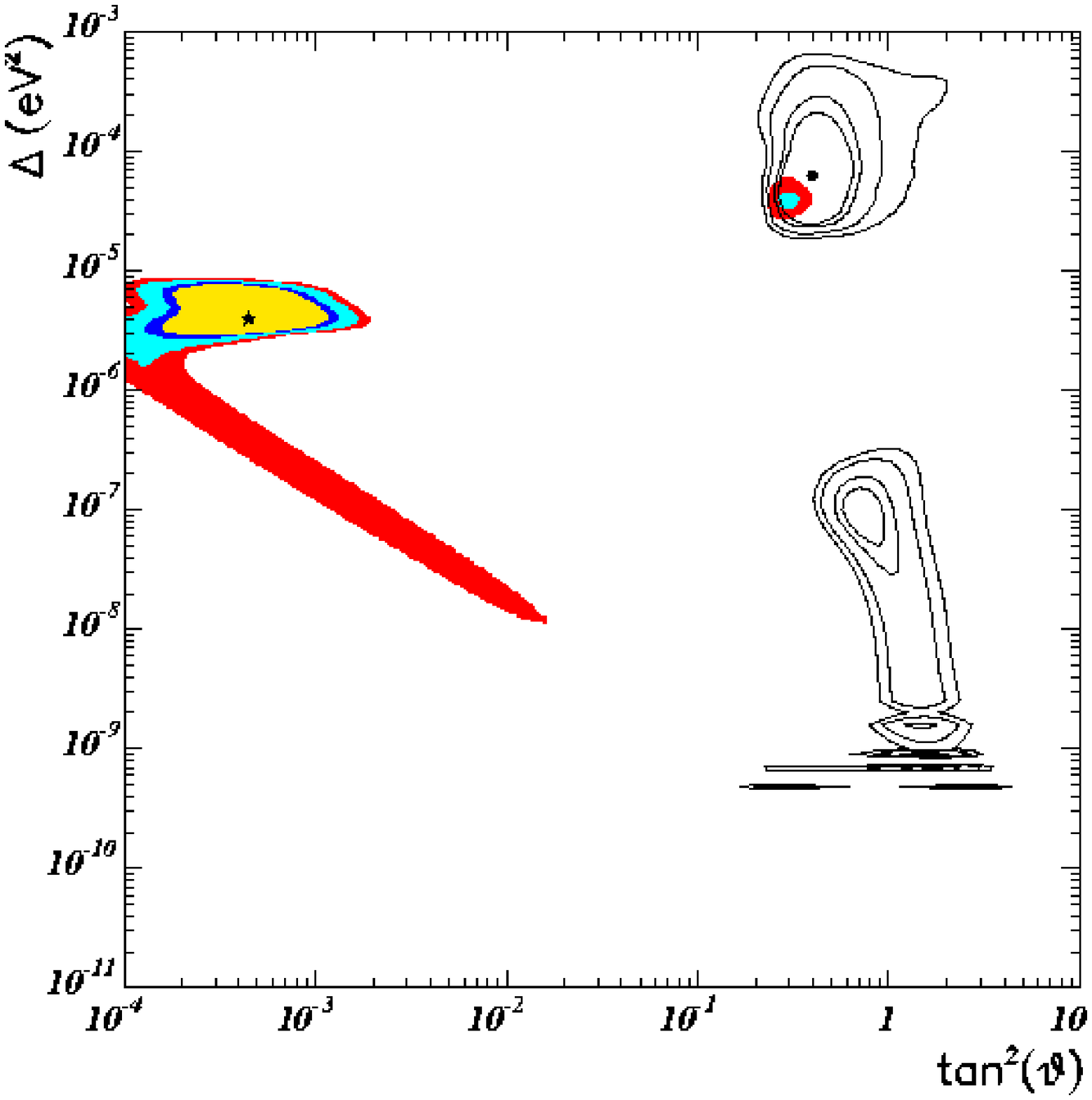,height=7.cm,width=9.cm,angle=0}
\epsfig{file=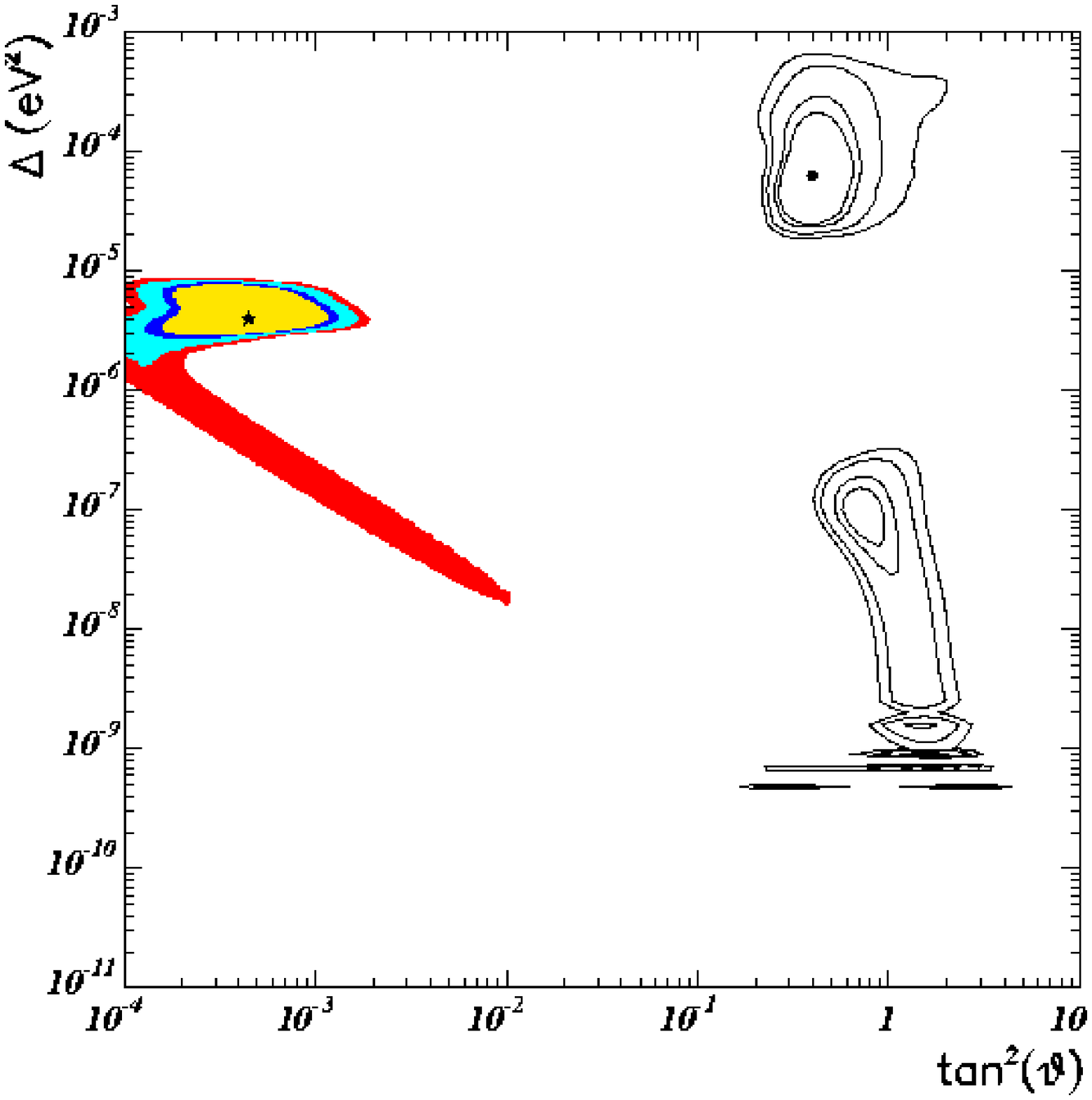,height=7.cm,width=9.cm,angle=0}
\caption{\label{3_16_tot}
  The 90, 95, 99 and 99.73\% C.L. contours of the combined fit of solar
and
  SN~1987A data (coloured/grey) together with the contours of the solar
  data alone (solid lines);
  for $\tau=\Eh/\Ee=1.4$ (top), $\tau=1.7$ (middle) and $\tau=2$
(bottom).
  All figures for $\Eb=3 \times 10^{53}$~erg and $\Ee=16$~MeV.}
\end{figure}
\end{center}

\begin{center}     
\begin{figure}
\hspace*{-0.4cm}
\epsfig{file=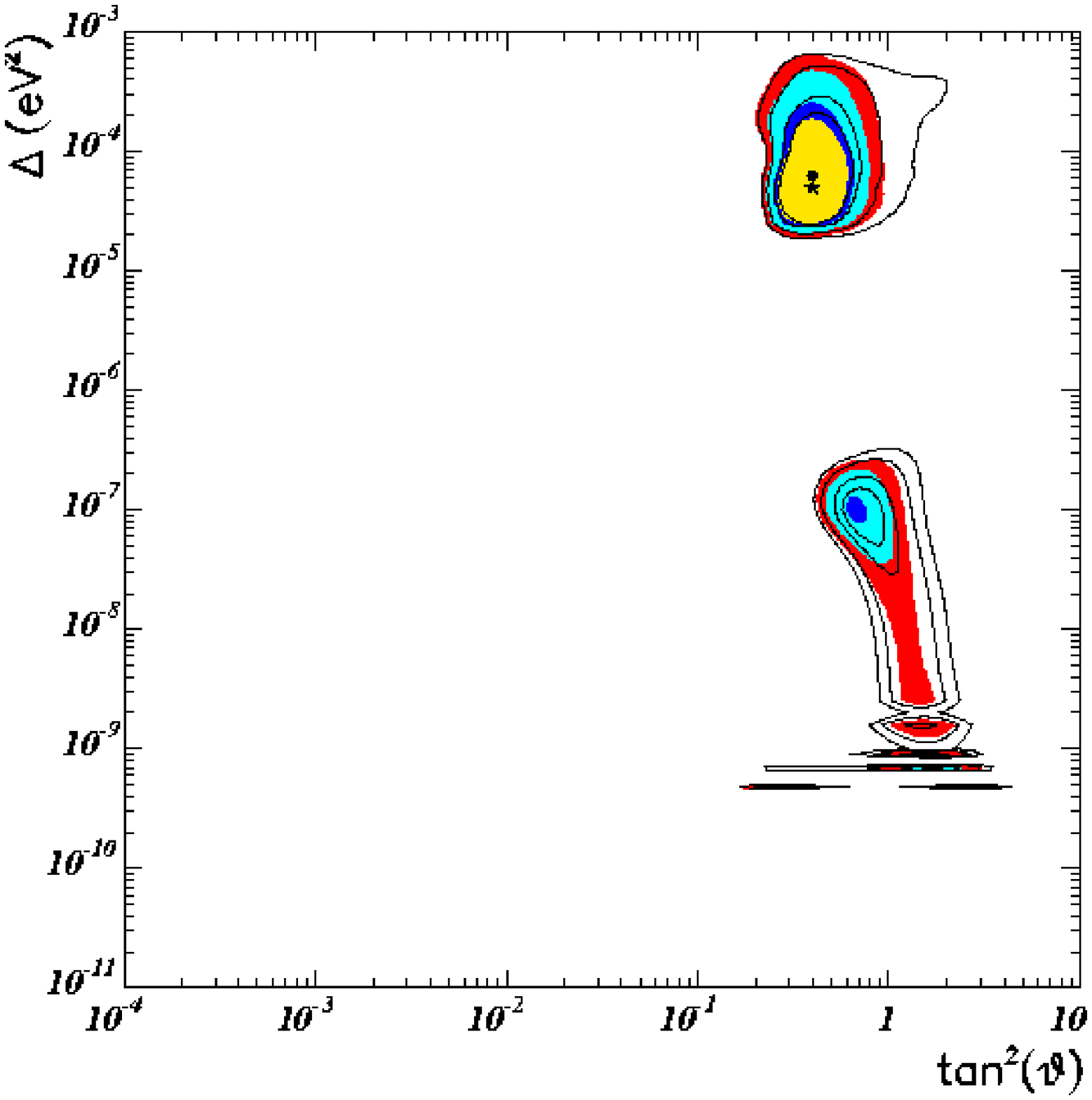,height=7.cm,width=9.cm,angle=0}
\epsfig{file=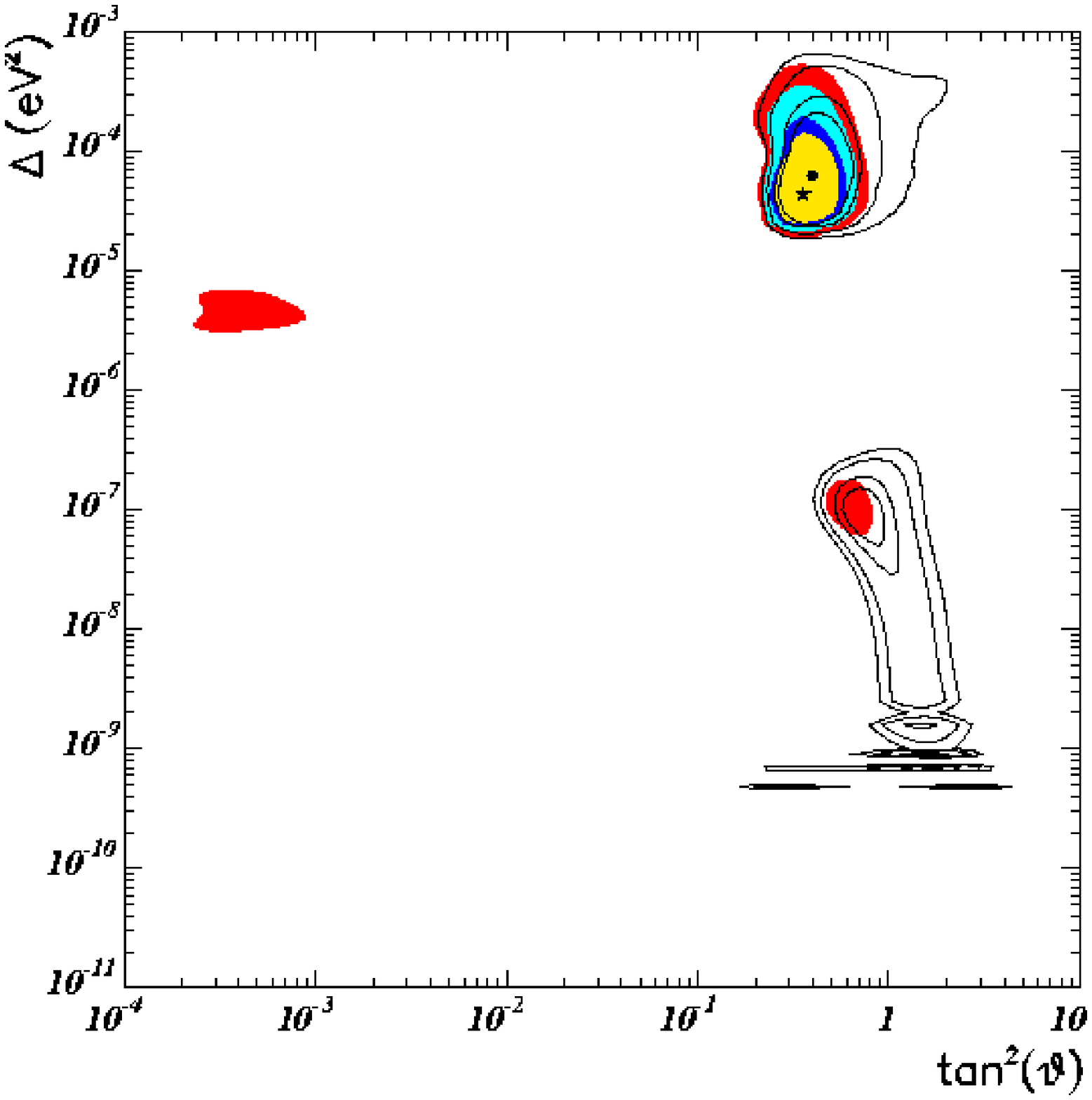,height=7.cm,width=9.cm,angle=0}
\epsfig{file=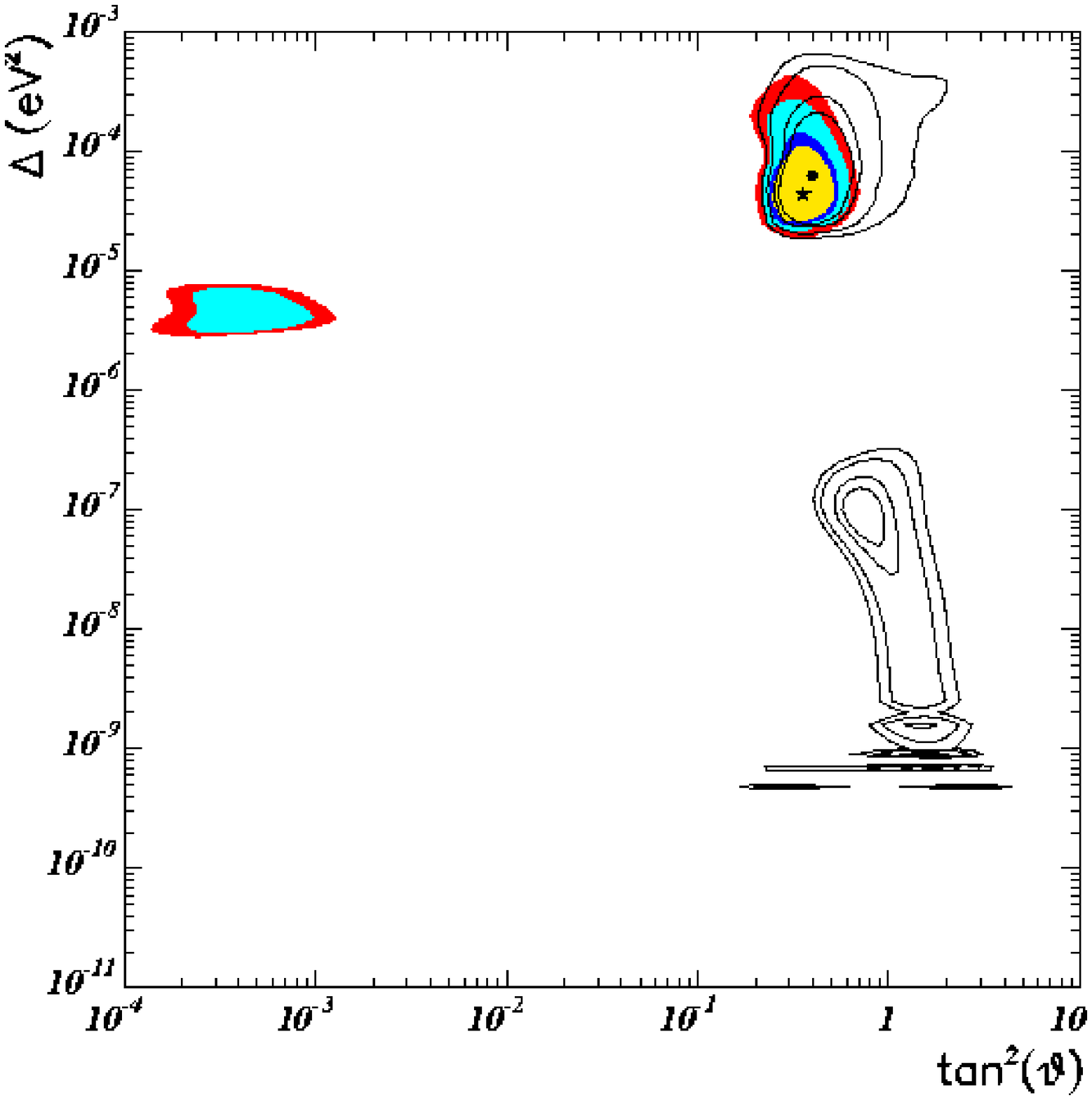,height=7.cm,width=9.cm,angle=0}
\caption{\label{1_14_tot}
  The 90, 95, 99 and 99.73\% C.L. contours of the combined fit of solar
and
  SN~1987A data (coloured/grey) together with the contours of the solar
  data alone (solid lines);
  for $\tau=\Eh/\Ee=1.4$ (top), $\tau=1.7$ (middle) and $\tau=2$
(bottom).
  All figures for $\Eb=1\times 10^{53}$~erg and $\Ee=14$~MeV.}
\end{figure}
\end{center}

\begin{center}      
\begin{figure}      
\hspace*{-0.4cm}
\epsfig{file=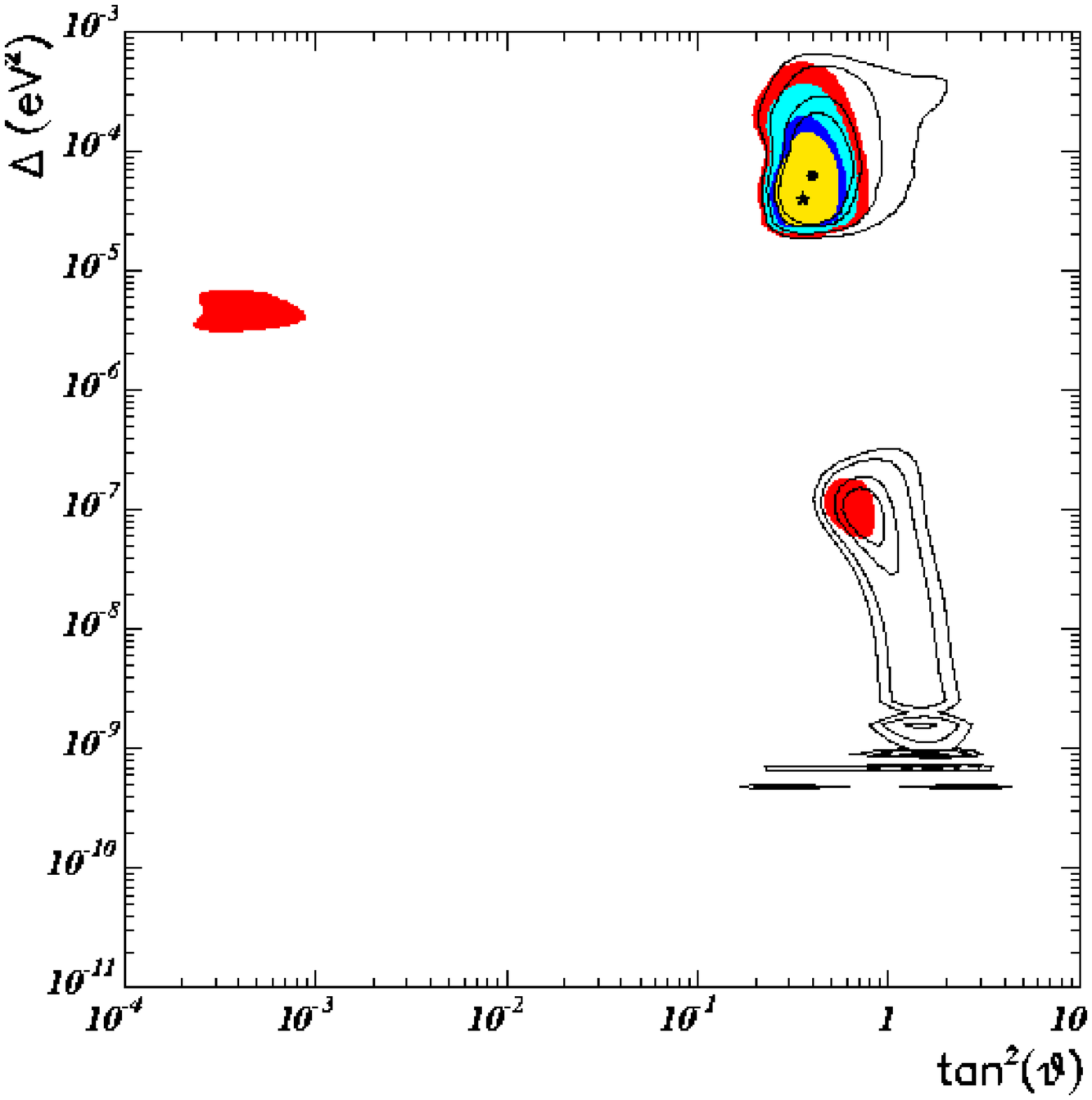,height=7.cm,width=9.cm,angle=0}
\epsfig{file=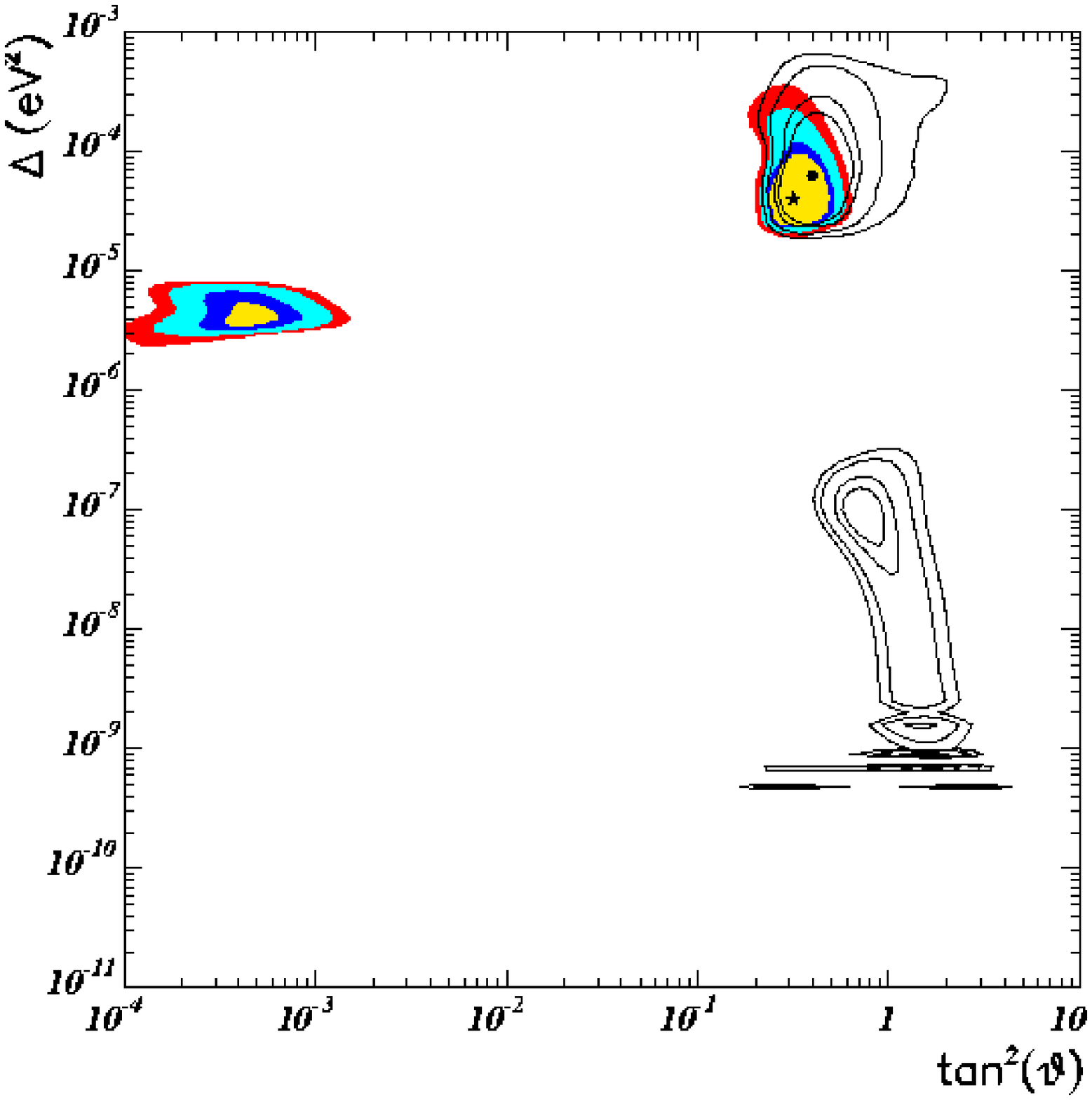,height=7.cm,width=9.cm,angle=0}
\epsfig{file=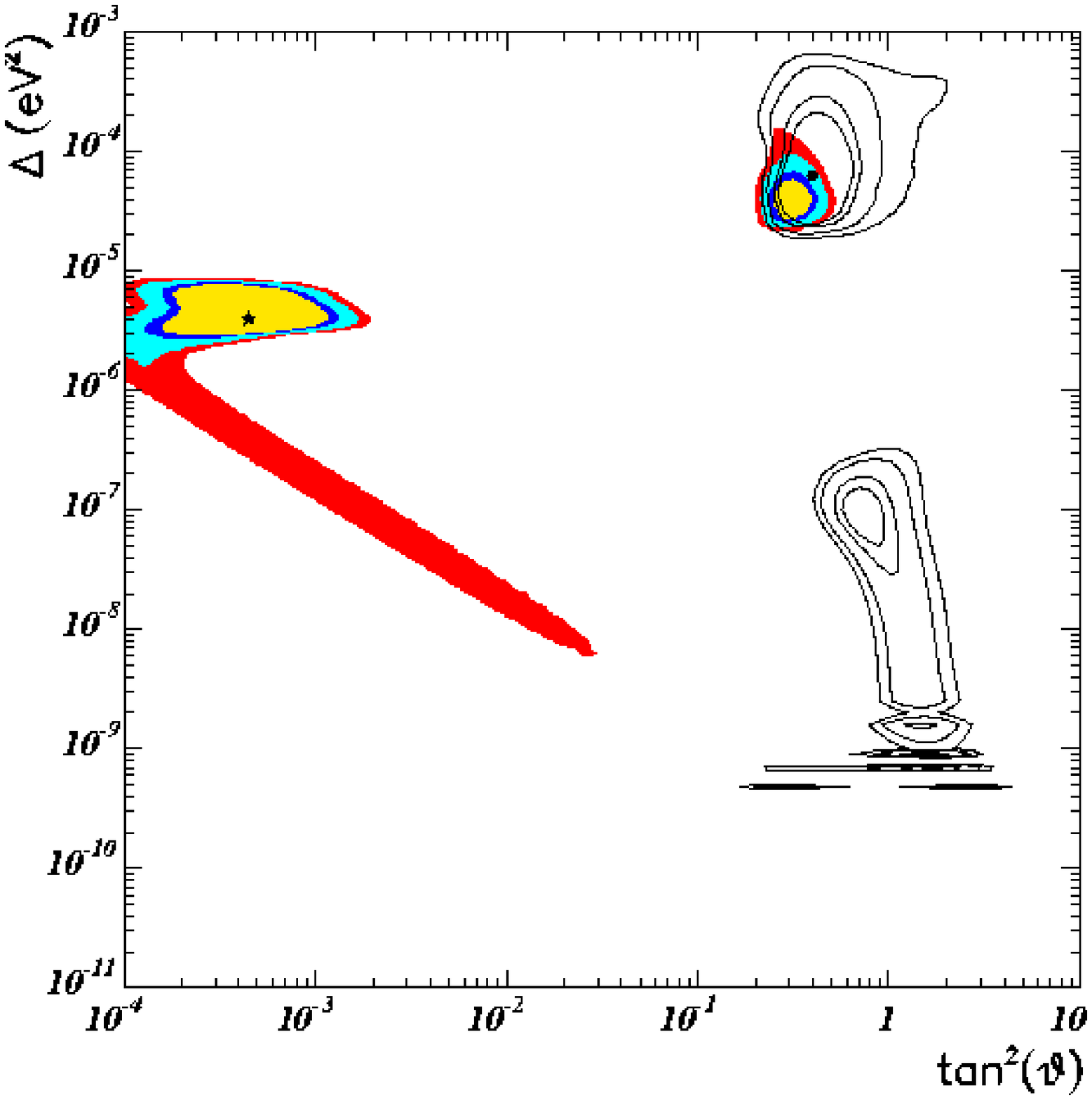,height=7.cm,width=9.cm,angle=0}
\caption{\label{3_12_tot}
  The 90, 95, 99 and 99.73\% C.L. contours of the combined fit of solar
and
  SN~1987A data (coloured/grey) together with the contours of the solar
  data alone (solid lines);
  for $\tau=\Eh/\Ee=1.4$ (top), $\tau=1.7$ (middle) and $\tau=2$
(bottom).
  All figures for $\Eb=3\times 10^{53}$~erg and $\Ee=12$~MeV.}
\end{figure}
\end{center}

In Figs.~\ref{3_14_tot}--\ref{3_12_tot} we show the 90, 95, 99 and
99.73\% C.L. contours of the combined fit of solar and SN~1987A data
for different astrophysical parameters. We also show the best-fit
point of the solar neutrino data alone (dot) and the best-fit point of
the combined data set (star). The values chosen in Fig.~\ref{3_14_tot}
correspond to a rather representative set of astrophysical parameters,
namely $\Eb=3\times 10^{53}$~erg and $\Ee=14$~MeV. In this case, the
impact
of the SN~1987A data on the standard solutions to the solar neutrino
problem is rather dramatic: the LOW-QVAC and VAC solutions disappear
for all assumed $\tau$ values.  We find that they are excluded at
99.98\%
and 99.99\% respectively, even for $\tau=1.4$.
Moreover the size of the LMA--MSW solution decreases, with increasing
$\tau$. The part of the LMA--MSW solution which is most stable against
the addition of the supernova data corresponds to the lowest $\Delta$
and
$\tan^2\t$ values, since these are favoured by Earth matter
regeneration effects. On the other hand the SMA--MSW region re-appears
extending, for increasing $\tau$, as a funnel towards the VAC solution
along the hypotenuse of the solar MSW triangle.  The combined best-fit
point moves from the LMA--MSW region for $\tau=1.4$ to the SMA--MSW
solution for $\tau=1.7$ and 2.  Comparing the C.L. contours for the
different solutions to the solar neutrino problem with the border
between the adiabatic and non-adiabatic region in Fig.~\ref{border1}, 
we find that the
SMA--MSW solution lies for anti-neutrinos always in the adiabatic
region, while it is in the non-adiabatic region for neutrinos (dotted
line).

In Fig.~\ref{1.5_12_tot}, we have chosen astrophysical parameters
corresponding to the lower limit of the range found in simulations,
$\Eb=1.5\times 10^{53}$~erg and $\Ee=12$~MeV.  Even in this case, the
vacuum type solutions to the solar neutrino problem are severely
disfavored, and exist very marginally only at the $3\sigma$ level for
$\tau=1.4$. In contrast, the LOW solution can still exist at 90\% level
for the most favorable choice of astrophysical parameters.  Finally
the LMA solution remains for all of the above three $\tau$ choices and,
most importantly, for such choices of astrophysical parameters the
best combined global fit-point remains within the LMA solution.

In order to analyse the stability of the LMA solution against larger
values of the astrophysical parameters, we have considered in
Fig.~\ref{3_16_tot}, the following case, $\Eb =3\times 10^{53}$ erg,
and $\Ee = 16$ MeV. It can be seen how even for this choice the
LMA-MSW solution is allowed at 99\% C.L unless $\Eh$ becomes
uncomfortably large ($\tau=2$), in which case no large mixing solutions
remains at $3\sigma$.

To elucidate more the importance of the different parameters, we show
in Figs.~\ref{1_14_tot} and \ref{3_12_tot} the confidence contours for
low $\Eb$ combined with an average value of $\Ee$ and vice versa.
Comparing the two set of figures, one recognizes that a low value of
$\Eb$ can be more important for the ``goodness-of-fit'' of the LMA--MSW
region than a low value of $\Ee$. In the case of small energy release,
$\Eb=1\times 10^{53}$~erg, the combined best-fit point remains even for
$\tau=2$ and $\Ee=14$~MeV in the LMA--MSW region.

\begin{center}      
\begin{figure} 
\epsfig{file=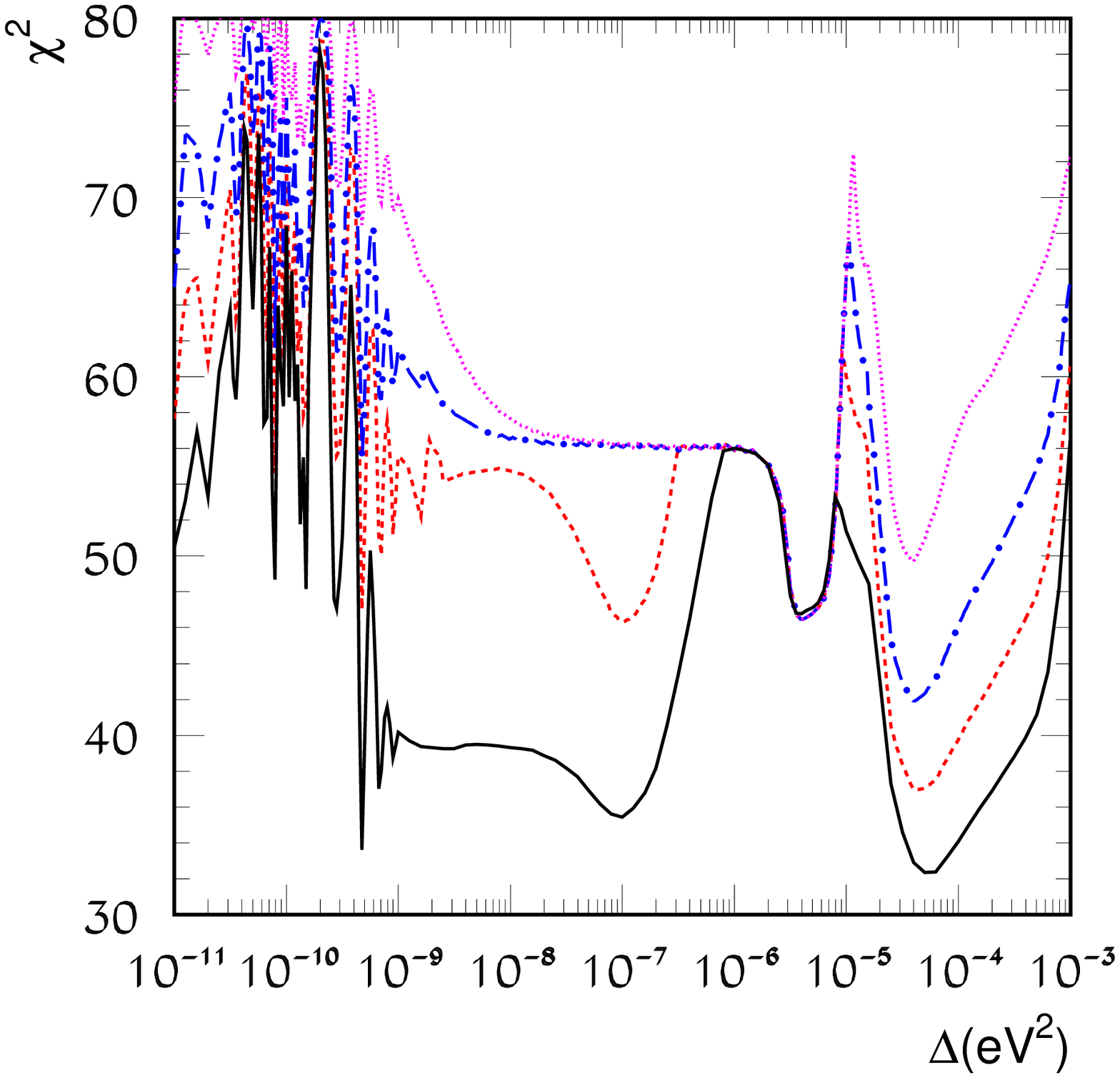,height=7.cm,width=9.cm,angle=0}
\caption{\label{combdm}
  The solid curve indicates the $\chi^2$ of the various
  oscillation solutions to the solar neutrino problem. The non-solid
  curves illustrate the effect of adding the SN~1987A data, which
  worsens the status of large mixing-type solutions.  See text for
  explanation.}
\end{figure}
\end{center}

\begin{center}       
  \begin{figure}
\epsfig{file=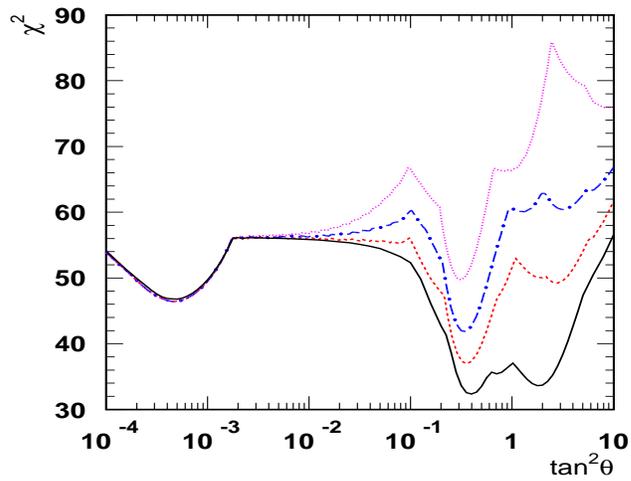,height=7.cm,width=9.cm,angle=0}
\caption{\label{combt2} Same as in Fig. \ref{combdm} but displayed
  with respect to $\t$; see text for explanation.}
\end{figure}
\end{center}

In Fig.~\ref{combdm} we illustrate in a global way the relative status
of various oscillation solutions to the solar neutrino problem and the
qualitative impact of adding the SN~1987A data. For each $\Delta$ value
we
have optimized the $\chi^2$ with respect to $\t$.  The
solid curve indicates the $\chi^2$ of the various
oscillation solutions to the solar neutrino problem. The non-solid
curves correspond to the case where the SN~1987A data are included.
The dash-dotted line is for $\Eb=3\times 10^{53}$~erg, $\tau=1.4$ and
$\Ee=14$~MeV.  The dashed line is for $\Eb=3\times 10^{53}$~erg,
$\tau=1.4$
and $\Ee=12$~MeV.  The dotted line is for $\Eb=3\times 10^{53}$~erg,
$\tau=1.7$ and $\Ee=14$~MeV. Here we have adjusted an arbitrary constant
which appears when combining the minimum likelihood-type SN~1987A
analysis with the solar $\chi^2$ data analysis in such a way that the
SMA
solution gets unaffected by the SN~1987A data. One notices that the
effect of adding SN~1987A data is always to worsen the status of
presently preferred oscillation-type solutions.  Within each such
curve one can compare the relative goodness of various solutions,
however different curves should not be qualitatively compared.
In Fig.~\ref{combt2} we repeat this procedure
after optimizing over $\Delta$ and displaying the result with respect to
the mixing parameter $\t$.

In contrast to earlier investigations, we conclude that the observed
neutrino signal of SN~1987A is compatible with the LMA--MSW solution,
especially for ``standard'' values of the neutrino energies together
with small values of $\Eb$~\footnote{A deviation of equipartition,
  i.e. a smaller luminosity of $\bar\nu_{\mu,\tau}$ than of $\bar\nu_e$,
 has a similar effect.}.  Smaller apparent values for $\Eb$ are
e.g. expected if SN~1987A was not spherically symmetric and the 
axis of maximal emission of neutrinos was not pointing to the Earth. 
We note also that, although the size of the
LMA--MSW region in the combined fit may be substantially reduced, the
position of its local best-fit point is rather stable: the inclusion
of the SN~1987A data drives the local best-fit point to only slightly
smaller values of $\t$ and $\Delta$.  While there is no significant
conflict between the solar and SN~1987A data for the LMA--MSW
solution, the case of the other large mixing solutions is different:
even the LOW solution which always gives a better fit than the
vacuum-type solutions requires a conspiracy of all astrophysical
parameters to fit well the combined data.

\section{Summary and Discussion}
\label{sec:summary}

We have discussed non-adiabatic neutrino oscillations in general
power-law potentials $A\propto x^n$. We found that the resonance point
coincides only for a linear profile with the point of maximal
violation of adiabaticity. We presented the correct boundary between
the adiabatic and non-adiabatic regime for all $\t$ and $n$.  As a new
method to calculate the crossing probability also in the non-resonant
regime we proposed a different splitting of $\ln P_c$ into a $\t$
independent part $\kappa_n$ and new correction functions $\G_n$ which
have
a simple integral representation for all $\t\in [0\!:\!\pi/2]$.

As an important application for supernova neutrinos, we considered the
case $n=-3$ in detail.  Comparing the different approximations used in
the literature with exact numerical results, we found that all of them
fail in some part of the $\tan^2\t$--$\Delta$ plane. While the use of
$\F_1=1$ together with $A\propto r^{-3}$ describes correctly the
crossing
probability for small mixing in the resonant region, the errors become
larger for larger mixing until this approximation fails completely in
the non-resonant region.  The correction function $\F_\infty$ used for
$n=-3$ describes quite accurately the most interesting region of large
mixing as well as the non-resonant region, but does not reproduce the
correct shape of the MSW triangle.  In contrast, the results using the
$\F_{-3}$ and $\G_{-3}$ function agree very well with the numerical
results for a $1/r^3$ profile 
in the resonant region and for all $\t$, respectively.

Performing a combined likelihood analysis of the observed neutrino
signal of SN~1987A and solar neutrino data including the recent SNO CC
measurement, we have found that the supernova data offer additional
discrimination power between the different solutions
of the solar neutrino puzzle.
Unless all relevant supernova parameters lie close
to their extreme values found in simulations, the status of the large
mixing solutions deteriorates, although the LMA solution may still
survive as the best fit of solar and SN~1987A data for acceptable
choices of astrophysical parameters. In particular, SN~1987A data generally
favour its part with smaller values of $\t$ and $\Delta$.
In contrast the vacuum or ``just-so'' solution is excluded and the LOW
solution
is significantly disfavoured
for most reasonable choices of astrophysics parameters.
The SMA--MSW solution is absent at about the $3\sigma$-level if solar
data only are included but may reappear once SN~1987A data are added,
due to the worsening of the large mixing type solutions.

Finally, we speculate about the possible impact of the
SN~1987A data on the solar neutrino problem in the near future.  If
the solution of the solar neutrino puzzle lies in the LMA--MSW region,
the KamLAND experiment~\cite{KamL} will determine
$\t$ and $\Delta$ accurately~\cite{Kaml_pred} (unless $\Delta >
2~10^{-4}\,{\rm eV}^2$).
However KamLAND cannot distinguish $\t$ from $\pi/2-\t$.
Both solar and SN~1987A data can make this distinction, and both favour
$\t< \pi/4$:
in the solar case because it gives a MSW resonance in $\nu_e$
oscillations;
in the SN~1987A case because it gives no MSW resonance in $\bar{\nu}_e$
oscillations.
  On the other hand, the knowledge of the
neutrino mixing parameters would change the emphasis of studies of the
SN~1987A neutrino signal from particle physics to astrophysics and
help to shed new light on the allowed range of the astrophysical
parameters.

Last but not least, let us mention that one should not forget a
fundamental difference between the SN~1987A and the solar fits. In the
solar case, a well-tested standard solar model exists, whose errors in
the various quantities from astro- and nuclear physics can be
estimated and are accounted for in the fit. In contrast
there is no ``standard model'' for type II supernovae and
therefore also no well-established average values and error estimates
for the relevant astrophysical parameters.
Nevertheless, experimental data about SN~1987A exist,
and it is worth studying their consequences.

\acknowledgments
We thank Pere Blay, Thomas Janka, Hiroshi Nunokawa, Carlos Pe{\~n}a
and Alexei Smirnov for discussions.
R.T. would like to thank the Theory Division at CERN for hospitality
and the Generalitat Valenciana for a fellowship.  This work was also
supported by Spanish DGICYT grant PB98-0693, by the European
Commission RTN network HPRN-CT-2000-00148, and by the European Science
Foundation network N.86.







\newpage

\section*{Addendum: Combined analysis of SN~1987A and solar data after
  SNO's NC and day/night results} 
\label{sec:addendum}

In this addendum, we update our analysis including the new NC and
day/night measurements of SNO~\cite{SNOnew} as well as the latest
Super-K~\cite{SuperKnew}, SAGE~\cite{SAGEnew} and GNO
results~\cite{GNOnew}. As a result
of the precise determination of the $^8$B flux by SNO together with
the absence of spectral distortions in Super-K, the solar data
disfavour now SMA-MSW, VAC, and the LOW solutions more strongly than
before.  
To show the impact of the new measurements, we have performed a new
combined $\chi^2$ analysis of the SN~1987A and the solar data, using
the same method as described in section~\ref{combinedanalysis}. The
$\chi^2_\odot$ were taken from Ref.~\cite{newsolar}, where also 
more details of this analysis can be found.

We consider again the astrophysical parameters ($\Eb$, $\Ee$, $\tau$
and implicitly the assumption of equipartition) as given {\em a
  priori}, and use only $\vartheta$ and $\Delta$ as fit parameters.
Since present simulations with improved treatment of neutrino
transport and microphysics (inclusion of energy transfer by recoils,
nucleon bremsstrahlung, and flavour-changing neutrino annihilations
$\nu_e\bar\nu_e\to\nu_{\mu,\tau}\bar\nu_{\mu,\tau}$) find much smaller
differences between the spectra of $\bar\nu_e$ and
$\bar\nu_{\mu,\tau}$, we use now only the lowest value for the ratio
$\tau$ of the $\bar\nu_{\mu,\tau}$ and $\bar\nu_e$ temperatures
considered previously, $\tau=1.4$~\cite{newtau}.
There are two main consequences of low $\tau$ values: First, the
weight of the SN~1987A data in the combined fit is reduced. Second,
deviations from equipartition which are naturally expected become more
important and should not be neglected in the future.

\begin{center}     
\begin{figure}
\hspace*{-0.4cm}
\epsfig{file=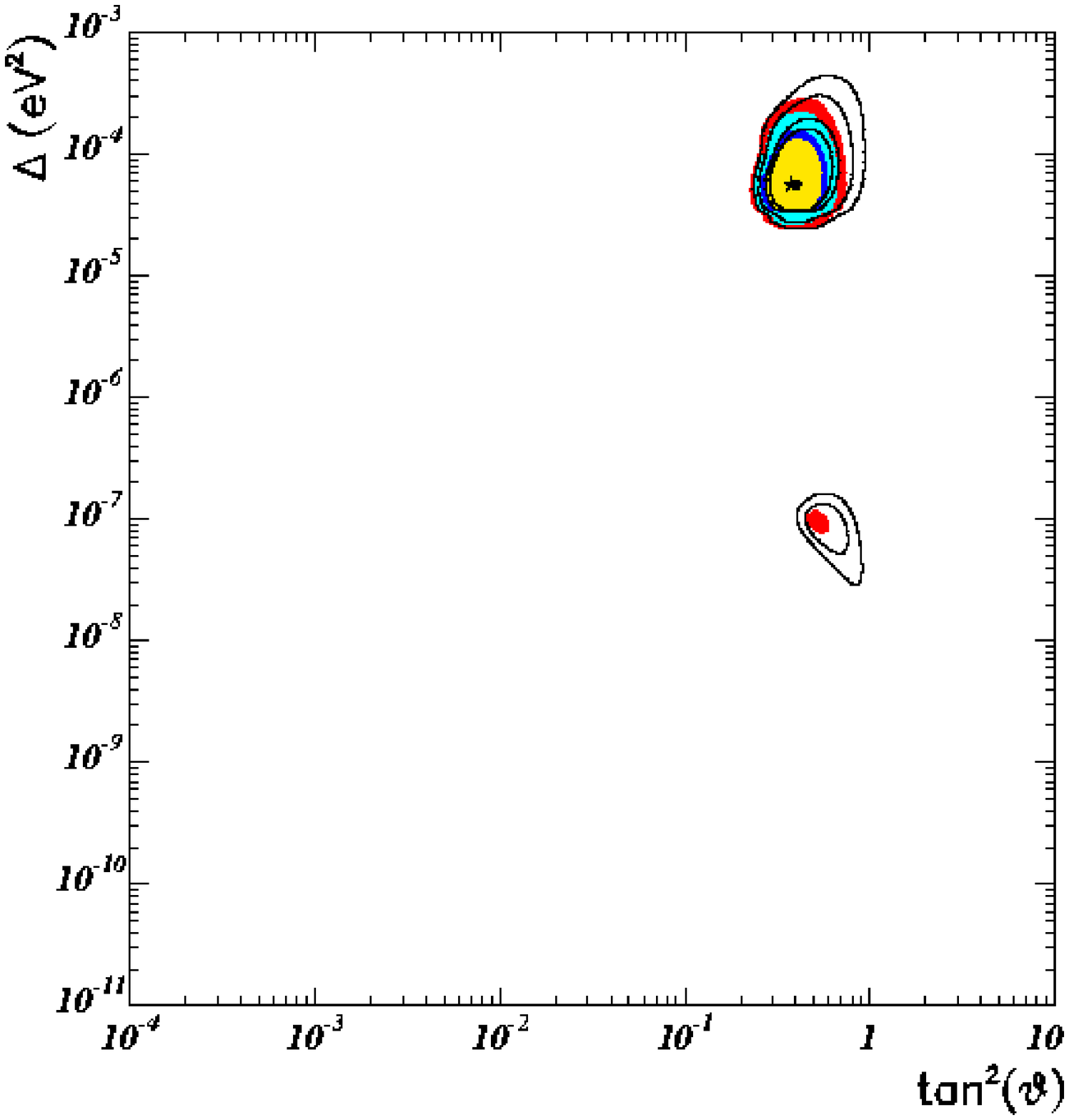,height=7.cm,width=9.cm,angle=0}
\epsfig{file=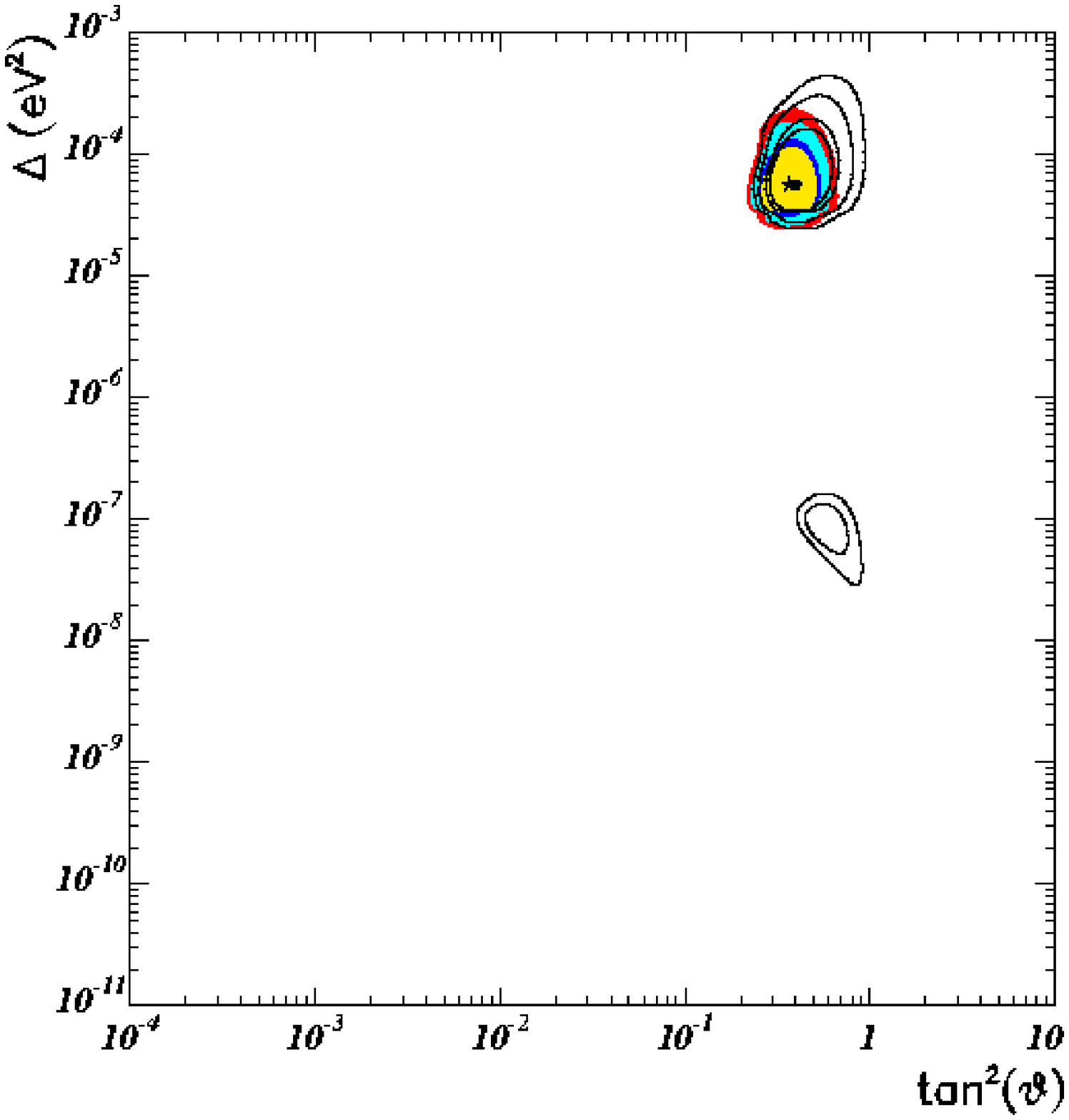,height=7.cm,width=9.cm,angle=0}
\epsfig{file=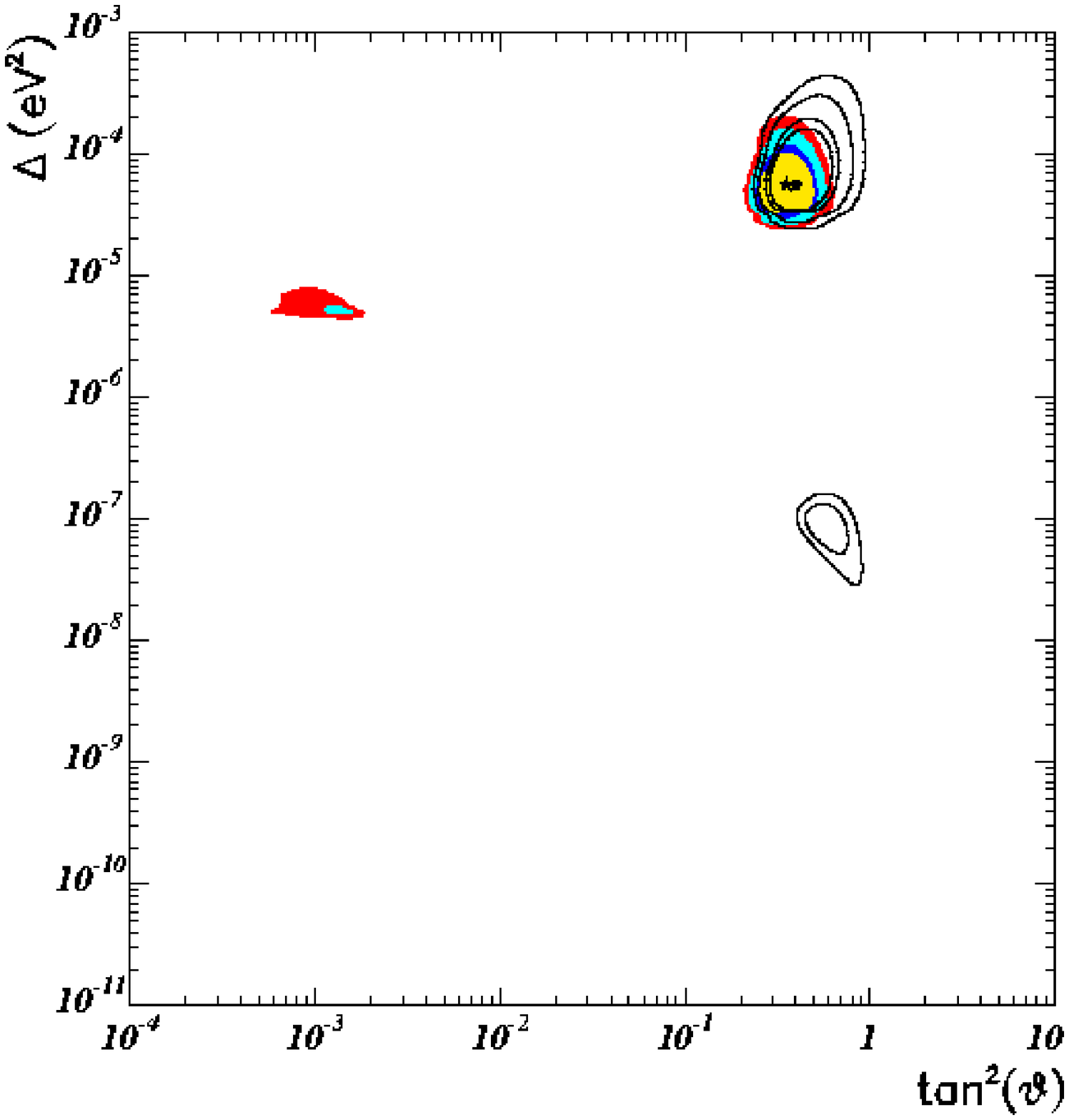,height=7.cm,width=9.cm,angle=0}
\caption{\label{new_figures}
  The 90, 95, 99 and 99.73\% C.L. contours of the combined fit of
  solar and SN~1987A data (coloured/grey) together with the contours
  of the solar data alone (solid lines); for $\Ee=12$~MeV (top),
  $\Ee=14$~MeV (middle) and $\Ee=16$~MeV (bottom).  All figures are
  for $\Eb=3\times 10^{53}$~erg and $\tau=\Eh/\Ee=1.4$.}
\end{figure}
\end{center}

In Fig.~\ref{new_figures}, we show the 90, 95, 99 and 99.73\% C.L.
contours of the combined fit for different $\bar\nu_e$ average
energies, $\Ee=12,~14$ and 16~MeV.  In all cases, we use $\Eb=3\times
10^{53}$ erg and $\tau=1.4$; also, we keep the assumption of
equipartition to allow a simple comparison with the previous figures.
The best-fit point of the combined data set lies now for all values of
$\Ee$ in the LMA-MSW region, very close to the solar best-fit point.
Both the new solar data and the small value of $\tau$ strengthen the
status of LMA-MSW as the leading solution to the solar neutrino
problem. 
Only the regions of the  LMA-MSW solution with large $\Delta$
or $\tan\t\gsim 1$ --- where Earth regeneration effects are less
important --- are inconsistent with the combined data set.
As in our previous analysis, we find that the other large
mixing solutions, LOW, quasi-VAC and VAC, are disfavoured by the
SN~1987A data.  The LOW solution appears for small values of $\Ee$,
but only at the 99.73\%~confidence level. On the other hand, the
SMA-MSW solution appears at 99\%~C.L.  if one goes to large values of
$\Ee$.

\begin{center}     
\begin{figure}
\hspace*{-0.4cm}
\epsfig{file=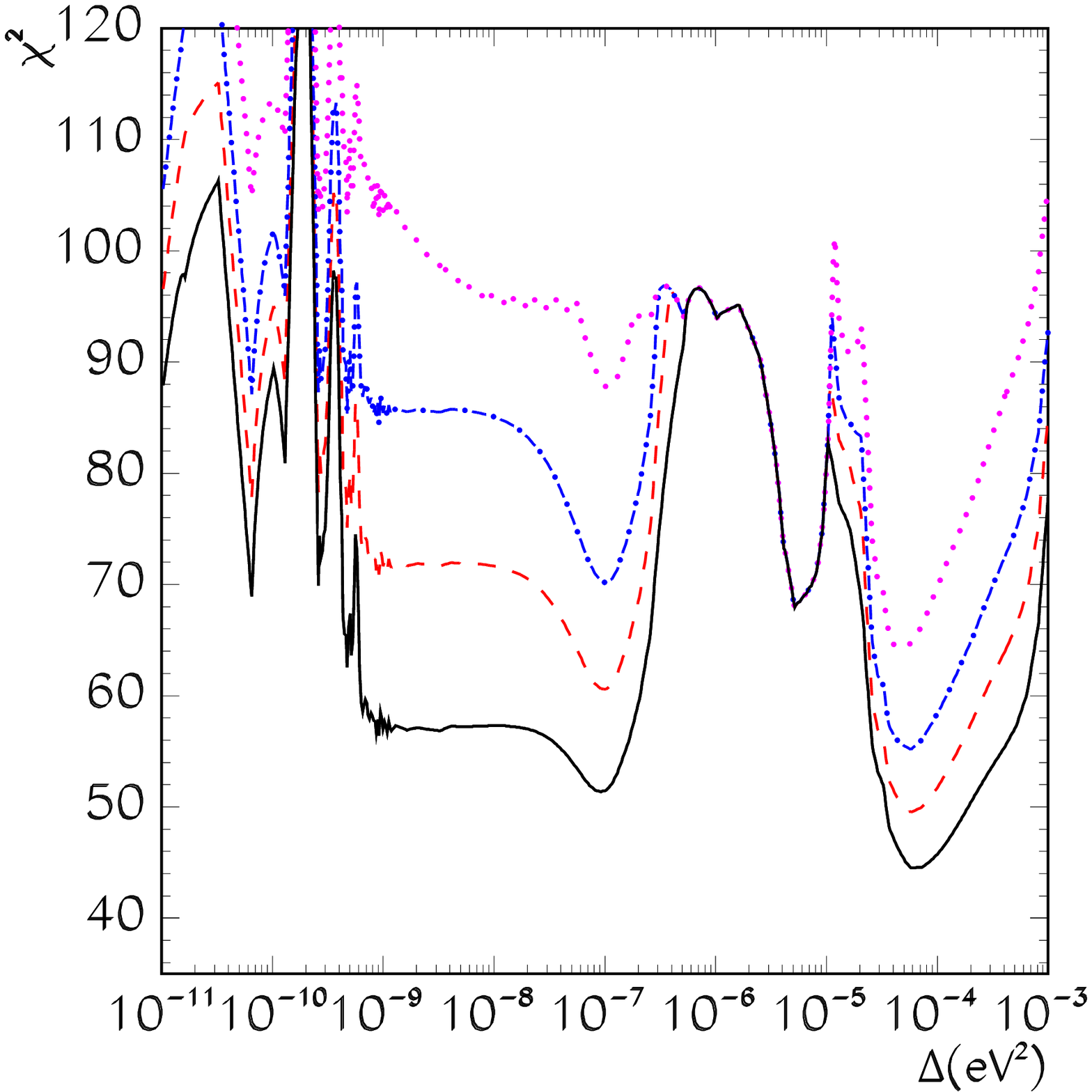,height=7.cm,width=9.cm,angle=0}
\epsfig{file=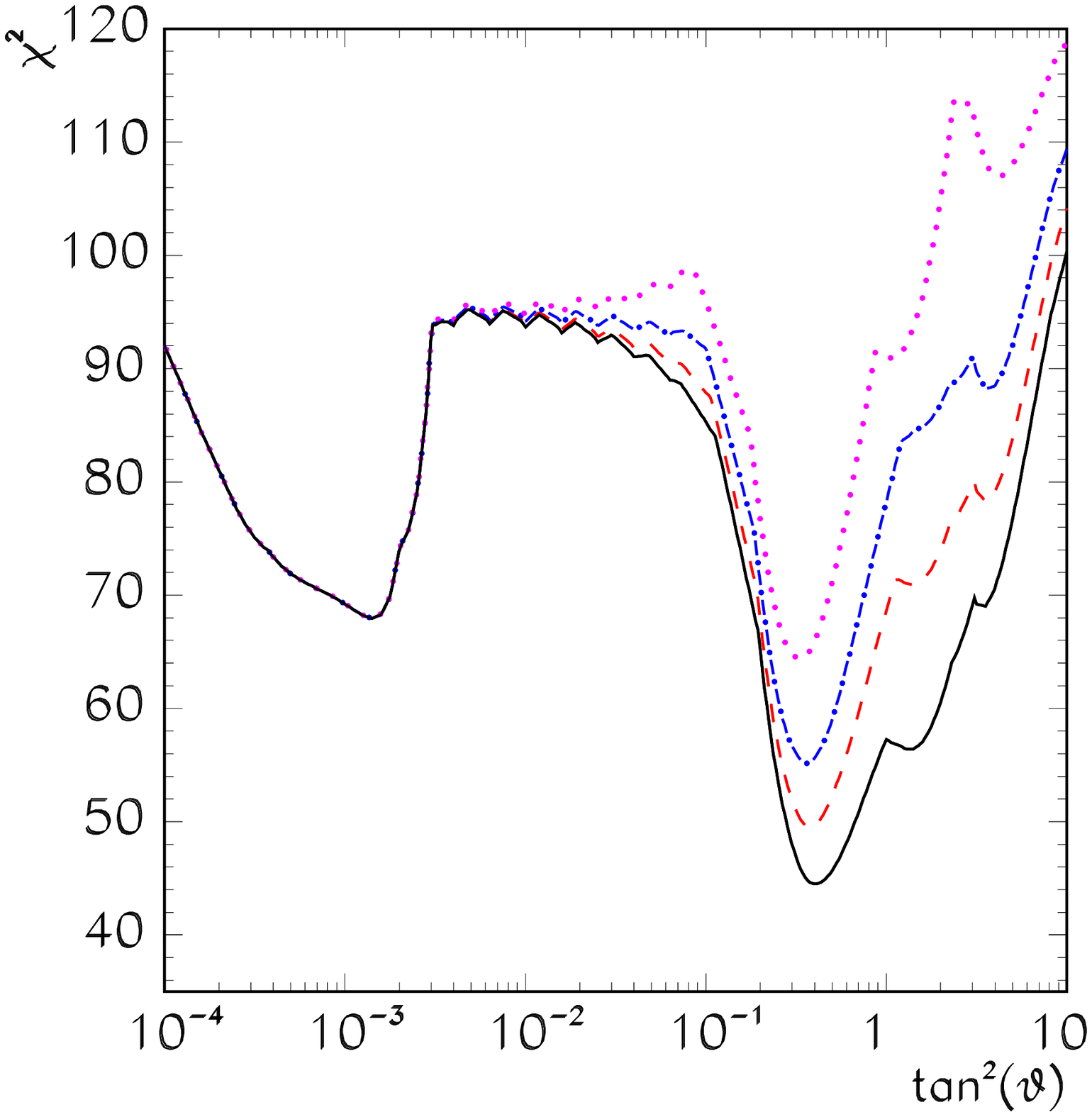,height=7.cm,width=9.cm,angle=0}
\vskip0.5cm
\caption{\label{new_figures2}
  The solid curve indicates the $\chi^2$ of the various
  oscillation solutions to the solar neutrino problem as a function of
  $\Delta$ (top) and $\tan^2\vartheta$ (bottom). The non-solid
  curves illustrate the effect of adding the SN~1987A data, which
  worsens the status of large mixing-type solutions.  See text for
  explanation.}
\end{figure}
\end{center}

As a concise way to illustrate these results, we show in
Fig.~\ref{new_figures2} the $\chi^2$ in terms of only one variable,
$\Delta$ or $\tan^2\vartheta$, respectively.  The dash-dotted line is
for $\tau=1.4$ and $\Ee=14$~MeV, the dashed line for $\tau=1.4$ and
$\Ee=12$~MeV, while the dotted line shows $\chi^2$ for
$\tau=1.7$ and $\Ee=14$~MeV.  
Finally, note that the arbitrary constant which appears when the
maximum-likelihood analysis of SN~1987A is combined with the solar
$\chi^2$ analysis has again been adjusted in such a way that
no oscillations (and consequently the SMA-MSW solution) are unaffected
by the SN~1987A data.

In summary, we find that the combined effect of current SN
simulations favoring low values of $\tau$, together with the large
$\chi^2_\odot$ difference between the LMA-MSW solution and its
competitors, result in a very comfortable position of the LMA-MSW
solution as the leading solution to the solar neutrino problem.

\acknowledgments
We thank Mathias Keil for useful discussions about the current status
of SN simulations. In Munich, this work was supported by the Deutsche
Forschungsgemeinschaft.  This work was also supported by Spanish
DGICYT grant PB98-0693, by the European Commission RTN network
HPRN-CT-2000-00148, and by the European Science Foundation network
N.86.

\newpage

\begin{thebibliography}{00}

\bibitem{review} For a comprehensive list of references see, e.g., the
  proceedings of the ``19th International Conference on Neutrino
  Physics and Astrophysics (Neutrino 2000)'', Nucl. Phys. B (Proc.
  Suppl.) {\bf 91} (2001).

\bibitem{Wo78} L.~Wolfenstein,
Phys.\ Rev.\ D {\bf 17}, 2369 (1978).

\bibitem{Mi85} S.P.~Mikheev and A.Yu.~Smirnov,
Sov.\ J.\ Nucl.\ Phys.\ {\bf 42}, 913 (1985);
Nuovo Cim.\ C {\bf 9}, 17 (1986).

\bibitem{cross}
W.C.~Haxton,
Phys.\ Rev.\ Lett.\ {\bf 57}, 1271 (1986);
S.J.~Parke,
{\it ibid.}, 1275 (1986).

\bibitem{LSZ}
L.~Landau, Phys. Z. Sowjetunion {\bf 2}, 46 (1932);
E.C.G. St{\"u}ckelberg, Helv. Phys. Acta {\bf 5}, 369 (1932);
C.~Zener,
Proc.\ Roy.\ Soc.\ Lond.\ A {\bf 137}, 696 (1932).


\bibitem{Ku89}
T.K.~Kuo and J.~Pantaleone,
Phys.\ Rev.\  {\bf D39}, 1930 (1989).

\bibitem{others}
C.~Sun,
Phys.\ Rev.\ {\bf D38}, 2908 (1988);
A.~Nicolaidis,
Phys.\ Lett.\ B {\bf 242}, 480 (1990).
M.M.~Guzzo, J.~Bellandi and V.M.~Aquino,
Phys.\ Rev.\ {\bf D49}, 1404 (1994).

\bibitem{l}
M.~Kachelrie{\ss} and R.~Tom{\`a}s,
Phys.\ Rev.\ D {\bf 64}, 073002 (2001).

\bibitem{solar} We use the data set corresponding to Fig.~6b
  of the updated version of P.~Creminelli, G.~Signorelli and A.~Strumia,
  ``Frequentist analyses of solar neutrino data,'' JHEP {\bf 0105}
  (2001) 052 found in hep-ph/0102234 which includes the SNO CC result.
  Note that we use confidence contours appropriate to two fit
  parameters, since we are working within a two--neutrino oscillation
  scenario. A three--neutrino analysis of neutrino
  data~\cite{Gonzalez-Garcia:2001sq} shows that this is a good
  approximation in view mainly of reactor bounds.

\bibitem{solar2}
V.D.~Barger, D.~Marfatia and K.~Whisnant,
hep-ph/0106207;
%
G.L.~Fogli, E.~Lisi, D.~Montanino and A.~Palazzo,
Phys.\ Rev.\ D {\bf 64}, 093007 (2001);
%
J.N.~Bahcall, M.C.~Gonzalez-Garcia and C.~Pe\~na-Garay,
JHEP {\bf 0108}, 014 (2001);
A.~Bandyopadhyay, S.~Choubey, S.~Goswami and K.~Kar,
Phys.\ Lett.\ B {\bf 519}, 83 (2001).


\bibitem{Mi87}
S.P.~Mikheev and A.Yu.~Smirnov,
Sov.\ Phys.\ JETP {\bf 65}, 230 (1987).

\bibitem{Fr00}
A.~Friedland,
Phys.\ Rev.\ {\bf D64}, 013008 (2001).

\bibitem{Pr90} A.P.~Prudnikov, Yu.A.~Brychokov and O.I.~Marichev, {\em
    Integrals and Series}, Gordon and Breach, New York 1990.

\bibitem{SN}
K.~Nomoto and M.~Hashimoto, Phys. Rept. {\bf 163}, 13 (1988);
H.A.~Bethe, Rev. Mod. Phys. {\bf 62}, 802 (1990).

\bibitem{snbounds}
J.~Arafune, M.~Fukugita, T.~Yanagida and M.~Yoshimura,
Phys.\ Lett.\  {\bf B194}, 477 (1987);
D.~Notzold,
Phys.\ Lett.\  {\bf B196}, 315 (1987);
H.~Minakata and H.~Nunokawa,
Phys.\ Rev.\  {\bf D38}, 3605 (1988);
T.K.~Kuo and J.~Pantaleone,
Phys.\ Rev.\  {\bf D37}, 298 (1988);
A.Y.~Smirnov, D.N.~Spergel and J.N.~Bahcall,
Phys.\ Rev.\  {\bf D49}, 1389 (1994);
A.S.~Dighe and A.Y.~Smirnov,
Phys.\ Rev.\ D {\bf 62}, 033007 (2000).

\bibitem{Gonzalez-Garcia:2001sq} M.C.~Gonzalez-Garcia, M.~Maltoni,
  C.~Pena-Garay and J.W.F.~Valle,
  Phys.\ Rev.\ D {\bf 63} (2001) 033005. 

\bibitem{Ka01}
M.~Kachelrie{\ss}, R.~Tom{\`a}s and J.W.F.~Valle,
JHEP{\bf 0101}, 030 (2001).

\bibitem{Wo}
S. Woosley, private communication. For more information see {\tt
  http://www.supersci.org/ } 

\bibitem{No}
T.~Shigeyama and K.~Nomoto, Astrophys. J. {\bf 360}, 242 (1990).

\bibitem{Jegerlehner:1996kx} T.J. Loredo, D.Q. Lamb, Ann. N. Y.
  Acad. Sci. {\bf 571} 601 (1989); 
B.~Jegerlehner, F.~Neubig and G.~Raffelt, 
Phys.\ Rev.\ D {\bf 54} (1996) 1194.

\bibitem{obs}
K.~Hirata {\it et al.}  [KAMIOKANDE-II Collaboration],
Phys.\ Rev.\ Lett.\  {\bf 58}, 1490 (1987);
R.M.~Bionta {\it et al.} [IMB Collaboration],
Phys.\ Rev.\ Lett.\  {\bf 58}, 1494 (1987).

\bibitem{hep-ex/0103033}
S.~Fukuda {\it et al.}  [Super-Kamiokande Collaboration],
Phys.\ Rev.\ Lett.\  {\bf 86}, 5656 (2001).

\bibitem{chooz}
M.~Apollonio {\it et al.}  [CHOOZ Collaboration],
Phys.\ Lett.\ B {\bf 466}, 415 (1999).

\bibitem{Minakata:2001rx}
H.~Minakata and H.~Nunokawa,
Phys.\ Lett.\ B {\bf 504} 301 (2001).

\bibitem{simu}
H.-T.~Janka, in Proc. {\em Frontier Objects in Astrophysics and
Particle Physics}, Vulcano 1992, eds. F. Giovaelli and G. Mannocchi;
A.~Burrows, T.~Young, P.~Pinto, R.~Eastman and T.~Thompson,
Astrophys.\ J.\ {\bf 539}, 865 (2000).


\bibitem{KamL}
P.~Alivisatos {\it et al.},
``KamLAND: A liquid scintillator anti-neutrino detector at the
Kamioka site,''
STANFORD-HEP-98-03.

\bibitem{Kaml_pred}
V.~Barger, D.~Marfatia and B.~P.~Wood,
Phys.\ Lett.\ B {\bf 498}, 53 (2001);
R.~Barbieri and A.~Strumia,
JHEP {\bf 0012}, 016 (2000);
H.~Murayama and A.~Pierce,
hep-ph/0012075;
A.~de Gouv\^ea and C.~Pe{\~n}a-Garay,
hep-ph/0107186.


\end{thebibliography}

\begin{thebibliography}{0}

\bibitem{SNOnew}
Q.~R.~Ahmad {\it et al.}  [SNO Collaboration],
arXiv:nucl-ex/0204008; 
Q.~R.~Ahmad {\it et al.}  [SNO Collaboration],
arXiv:nucl-ex/0204009.

\bibitem{SuperKnew}
M.~B.~Smy,
arXiv:hep-ex/0202020.

\bibitem{SAGEnew}
J.~N.~Abdurashitov {\it et al.}  [SAGE Collaboration],
arXiv:astro-ph/0204245.

\bibitem{GNOnew}
T.~Kirsten, talk at ``XXth International Conference on Neutrino Physics
and Astro\-physics'', Munich 2002.

\bibitem{newsolar}
P.~Creminelli, G.~Signorelli and A.~Strumia,
JHEP {\bf 0105} (2001) 052
[arXiv:hep-ph/0102234]. Version 3:
addendum, discussing the implications of the NC and day/night SNO data.

\bibitem{newtau}
G.~G.~Raffelt,
Nucl.\ Phys.\ Proc.\ Suppl.\  {\bf 110} (2002) 254
[arXiv:hep-ph/0201099];
R.~Buras, H.~T.~Janka, M.~T.~Keil, G.~G.~Raffelt and M.~Rampp,
arXiv:astro-ph/0205006;
M.~T.~Keil, G.~G.~Raffelt H.~T.~Janka,
``Monte Carlo study of supernova neutrino spectra formation''
(in preparation). In the last work, values as low as $\tau\ap 1.1-1.2$
have been found.

\end{thebibliography}
\end{document}